\documentclass[pre,showpacs,preprint]{revtex4-1}
\usepackage{amsmath,amssymb}
\usepackage{rotating}
\usepackage[utf8]{inputenc}
\usepackage{graphicx} % Include figure files
\usepackage{epstopdf}

\begin{document}

\title{
Evidence for multiple
Liquid-liquid phase transitions in carbon, and the
 Friedel-ordering of its liquid state.
}

\normalsize

\author
{M.W.C. Dharma-wardana}
\email[Email address:\ ]{chandre.dharma-wardana@nrc-cnrc.gc.ca}
\author
{Dennis D. Klug.}
\affiliation{
National Research Council of Canada, Ottawa, Canada, K1A 0R6
}

\date{\today}
\begin{abstract}
Carbon, the fourth most abundant element in the Universe forms a metallic
fluid with transient covalent bonds on melting. Its liquid-liquid phase
transitions, intensely sought using simulations had been elusive. Here we
use density functional theory (DFT)  simulations with up to 108 atoms
 using molecular dynamics,  as well as one-atom DFT  as implemented
in the neutral pseudo-atom method where  multi-atom effects are treated by
 ion-ion correlation functionals. Both methods use electron-electron
 exchange correlation functionals for electron many-body effects. Here
 we show using  both methods, that liquid carbon displays multiple
liquid-liquid transitions linked to changes in coordination number in the density
 range 3 g/cm$^3$, to $\sim$ 6 g/cm$^3$  when a coordination
 number of 12  is reached. The transitions disappear by 4 eV in temperature.
The calculated pressures and transition densities are shown to be sensitive to
 the exchange-correlation functionals used. Significantly, we find that
 a simple metallic model yields the structure factors and thermodynamics
 with  quantitative accuracy, without invoking any covalent-bonding features.
 The ion-ion structure factor for these densities and temperatures  is found to
 have a subpeak tied to twice the Fermi wavevector, constraining the fluid in
 momentum space. The dominant Friedel oscillations forming the pair
 interactions correlate the ions and drive  the  multiple liquid-liquid
  phase transitions. Our results suggest that liquid carbon typifies
  a  class of fluids whose  structure is ordered by the long-ranged
 Friedel  oscillations  in the pair-potentials.
 These  results are critical to  terrestrial  and astrophysical studies,
 inertial  fusion using carbon drivers,  refined shock experiments, and  in
 seeking new carbon-based  materials.
\end{abstract}

\maketitle

\section{Introduction} The physics and chemistry of carbon
 is central to the  evolution of stars, exoplanets as well as
the earth, comets  even in the context of interstellar dust.
Hull {\it et al}~\cite{Hull20} have recently emphasized the importance of
 studying $l$-carbon in many chemical physics contexts,
while Lazicki {\it et al}~\cite{Lazicki21} find persisting
 diamond structures even at terapascal pressures.
Carbon plays crucial roles in many industrial processes, and in
frontier research into new carbon materials. It is used for
 encapsulating deuterium in inertial laser fusion. Its properties
 in the regime of warm-dense matter have been intensely studied.
Intriguingly, although metallic, liquid
carbon ($l$-carbon) shows covalent bonding characteristics and most theoretical
studies have attempted a detailed account of such interactions
that contain directional, orientational and other poly-atomic
molecular features, using expensive
 computations involving hundreds of atoms
~\cite{galli89,Benedict14,Vorberger20}.

 While phase transitions in solids involve  changes in symmetry or some
 `order parameter' of a  property like magnetism, liquid-liquid phase
 transitions (LPTs) typically show
 a discontinuity  or divergence in a thermodynamic property like the pressure,
 or the compressibility. Such LPTs may be caused by changes
 in short-ranged bonding effects, long-ranged non-directional
many-body  correlations or `volume forces'.

Carbon~\cite{galli89,DWP-Carb90} silicon, and germanium etc.,
 are  `tetrahedral fluids' that are  metallic with four free electrons
 per ion in  the range of densities studied here.
Supercooled liquid silicon ($l$-Si) undergoes a subtle structural liquid-liquid
transition~\cite{GaneshSiLPPT-09}. Simulations of $l$-Si in thermal equilibrium
show other LPTs as  well~\cite{cdwSi20}.
At higher densities, in analogy with LPTs found in
 alkali metals~\cite{Zong21}, free electron phases, electride etc.,
 may  also exist~\cite{cdw-Carbon10E6-21}. Although a transient
 covalent-bonding  picture  is popular in interpreting
results from density functional theory (DFT) and molecular dynamics (MD)
 simulations, here we show that a simple metallic picture
quantitatively accounts for  the structure and
thermodynamics  of $l$-carbon,  just as for $l$-Si. The metallic picture
 treats long-ranged Friedel oscillations and ionic interactions accurately,
 exposing hitherto elusive LPTs generic to this class of fluids.

LPTs in carbon were discussed  by
van Thiel {\it et al}~\cite{vanThiel93}, while  Glosli  {\it et al}~\cite{glosli99}
used the multi-center empirical `Brenner' potential and
found a single LPT involving changes in bonding and coordination.
However, a DFT-MD study using the Perdew-Burke-Ernzerhof (PBE)
exchange-correlation (xc) functional~\cite{PBE96}
found no such LPT~\cite{WuLPT02}. Kraus {\it et al}~\cite{kraus13} also found
empirical potentials to be unreliable and used $N$-atom
density functional calculations coupled to MD simulations in
their laser-shock experiments on carbon.

Kraus {\it et al} state that
``Simple potentials from fluid
theory like that of Lennard and Jones fail as well. We have
also tested combinations of screened Coulomb interactions
and Lennard-Jones-type interactions. All of these calculations
 always predict a first peak around $k$=4.3 \AA$^{-1}$
 or a correlation length of 1.45 \AA \ which is related to the mean
density of the sample. However, the measured data do
not support such a structure factor. On the other hand, the
observed structure factors agree very well with results from
DFT-MD simulations. Such simulations include the full
quantum nature of the electrons and, thus, go far beyond
simple pair interactions between the nuclei. In particular,
they are able to describe short-time bonding''.

While {\it ad hoc} simple `screened Coulomb interactions'
 do fail~\cite{Stanek21},  we show in this study (see
also Ref.~\cite{DWP-Carb90}) that properly
 constructed pseudopotentials and pair-potentials based on
linearly screened Coulomb interactions built on rigorous
first principles methods give  quantitative agreement with
multi-center DFT-MD calculations for $l$-carbon, {\it without}
the need for multi-center potentials.

Here we  use two independent DFT  methods to establish hither-to unexpected
multiple LPTs linked to changes in coordination number at the densities studied.
Unlike in previous studies, we have examined the compressibility of liquid
carbon on very fine grids of densities along several isotherms to expose
the presence of discontinuities in the compressibility, while similar
studies for $l$-aluminum in a corresponding density range show no such
discontinuities, as expected for aluminum.

 Since the melting point of diamond is $\sim 0.5$ eV in
 energy units of temperature $T$ used here, we
study the range $T= 1-10$ eV. Such systems are conveniently
experimentally studied  using short-pulse lasers~\cite{kraus13,Lazicki21}.
We show that $l$-carbon is typical of a class of fluids (e.g., C, Si, Ge)
whose structure is determined by strong electron scattering across the
 Fermi surface, with a momentum transfer of 2$k_F$, where $k_F$
is the Fermi wavevector.

While a fluid cannot support a static charge density
wave, the short-ranged structure of the fluid can get modified. Here it
produces a peak at 2$k_F$ by splitting the main peak in the structure factor
$S(k)$. This mechanism creates LPTs not found in other metallic liquids
 like $l$-aluminum where the compressibility shows no discontinuities
 in the relevant density range. This suggests a new class of
 momentum-constrained {\it Friedel ordered} fluids associated
 with strong 2$k_F$ scattering.  Highly refined shock-wave
experiments may expose these subtle LPTs.
\section{Density functional methods used in the study}
 The DFT methods used here are discussed
briefly  in the Appendix.
The DFT-MD method implemented in
 VASP~\cite{VASP} or ABINIT~\cite{ABINIT} uses $N$ atoms in the simulation
cell, typically with $N\sim 100-500$. It is referred to as
`Quantum MD' (QMD) here, and was used recently to study
 $l$-carbon by Vorberger {\it et al},~\cite{Vorberger20} where the
 Perdew-Burke- Ernzerhof (PBE)~\cite{PBE96}
xc-functional at $T$=0 was used to treat electron-electron
many-body effects.
Such multi-center densities are found to be sensitive to the xc-functional
used~\cite{Remsing17}. Thus results
for $l$-Si from the PBE functional differed
 significantly from those of the `strongly constrained and appropriately
 normed' (SCAN) functional~\cite{SCAN13}. In our work we have used
 the SCAN functional for $N$-atom QMD carbon simulations via the
VASP code in this study. The sensitivity to the number of atoms $N$
used in the simulation is investigated using QMD
calculations for $N=64$ and 108.

The very heavy computational effort in $N$-atom QMD approaches
can be removed while still retaining the  power and rigour of
DFT in a  properly formulated full one-body DFT where $N=1$.
According to Hohenberg and Kohn, {\it one-body}  densities
 of electrons and ions completely determine the thermodynamics of
 {\it any} arbitrary electron-ion system.
Hence, in principle, $N$-atom DFT with $N$=1 should provide accurate
results if multi-ion effects could be handled  by a suitable ion-ion
xc-functional in addition to the usual electron-electron xc-functional.
Such a one-atom DFT is utilized in the neutral pseudoatom
 (NPA) method used
 here~\cite{cdwSi20,eos95}. Its accuracy has been established by
many previous calculations~\cite{cdwSi20,Hungary16} of static as well as
dynamic properties~\cite{DSF18}.
The many-ion and many-electron effects are included in one-atom DFT via
 exchange-correlation (xc) functionals for electrons as well as for
ions (see Appendix).

\begin{figure}[t]
\includegraphics[width=.98\columnwidth]{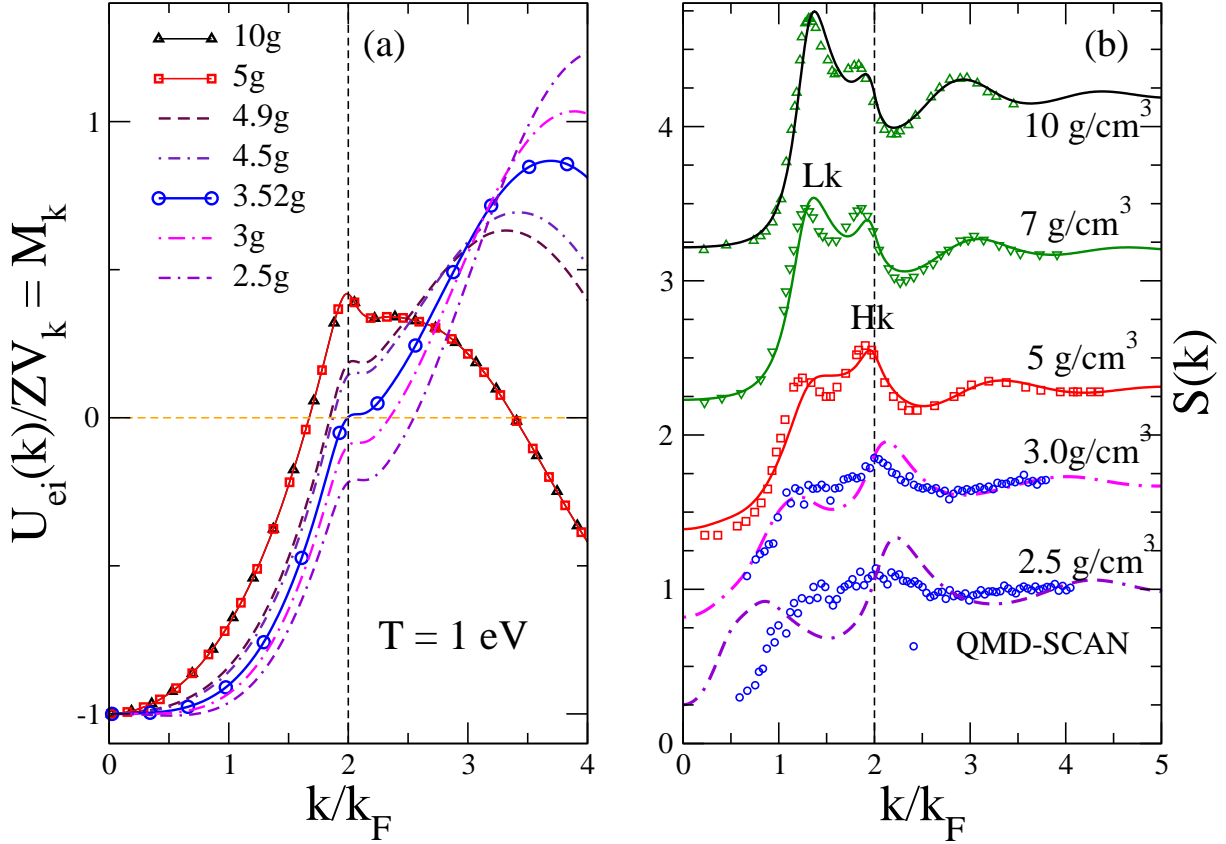}
\caption
{(Colour online)  Pseudopotentials and structure factors.
(a) The LR-NPA pseudopotential
$U_{ei}(k)$ for $l$-carbon at 1 eV for
2.5 g/cm$^3 \le \bar{\rho} \le$ 10 g/cm$^3$. The $U_{ei}$ is invariant
between 5-10 g/cm$^3$; a sharp change
 occurs  below 5 g/cm$^3$. The
$U_{ei}(k=2k_F)\to 0$ at $\rho_{\rm D}\simeq 3.5$ g/cm$^3$ (line with circles).
(b) shows typical $S(k)$ using the LR-$U_{ei}(k)$
 and the electron
response function $\chi_{ee}(k,T,\bar{n})$. The unconnected symbols
for $\bar{\rho}\ge 5$g/cm$^3$ are the QMD $S(k)$ of Ref.~\cite{Vorberger20}
 using the PBE functional.
 The smooth curves are NPA calculations. The high-$k$ subpeak
 (marked Hk) falls on 2$k_F$ for $\bar{\rho}\ge3$ g/cm$^3$.
We compare the LR-NPA $S(k)$ and QMD-SCAN $S(k)$ at 3 g/cm$^3$,
and 2.5 g/cm$^3$  showing the failure of LR below 3 g/cm$^3$.
%\label{Fig.1}
\label{UeiSk.fig}
}
\end{figure}

\section{The pseudopotentials and pair-potentials generated from the NPA}
The all-electron NPA calculation  self-consistently
 generates the Kohn-Sham one-body electron charge distribution
 (see Appendix) around
a carbon nucleus placed in the appropriate environment in the fluid.
This charge distribution is used to construct a pseudopotential
which is then employed to construct a pair-potential. If a
metallic model with strong screening is assumed, the above steps
can be carried out within linear response (LR) theory. Let
the free electron density around the ion of average charge $\bar{Z}$
be denoted by  $n_f(r)$, while the average (uniform) free electron density in
the fluid is denoted by $\bar{n}$. In our case $\bar{Z}=4$.
 We use the free-electron charge pileup
$\Delta n_f(r)=n_f(r)-\bar{n}$ and its Fourier transform $\Delta n_f(k)$
to construct the electron-ion  pseudopotential, in Hartree atomic units,
\begin{eqnarray}
\label{uei.eqn}
U_{ei}(k)&=&\Delta n_f(k)/\chi(k,r_s,T)\\
         &=&-\bar{Z}V_kM_k,\; V_k=4\pi/k^2.
\end{eqnarray}
Here $\chi(k,r_s,T)$, abbreviated to $\chi(k)$  is the fully
interacting finite-temperature
response function of the uniform electron fluid at the
density $\bar{n}$ associated with the electron Wigner-Seitz
radius $r_s$. The details of the calculation of $\chi(k)$
as well as the limitations of this pseudopotential are discussed
 in the Appendix.
Here we note that a full electron non-linear DFT calculation has been
applied to the electron-nuclear interaction in obtaining $\Delta n_f(r)$,
and LR is used only in defining a pseudopotential. Thus the pseudopotential
incorporates the  non-linear aspects of the Kohn-Sham calculation
contained in the NPA model.

The pseudopotential $U_{ei}(k)$ is re-expressed in terms
of the point-ion potential $\bar{Z}V_k$ and the form factor $M_k$
which is unity only for a point ion. This is a simple local ($s$-wave)
potential and we have found it adequate for quantitatively reproducing
results obtained from complex $N$-atom QMD using advanced non-local
pseudopotentials for a variety of materials. The resulting
 pseudopotentials are shown in panel (a) of figure~\ref{UeiSk.fig}.

The pair-potential within linear response that results from
 Eq.~\ref{uei.eqn} is, in Hartree atomic units:
\begin{equation}
\label{pair.eqn}
V_{ii}(k)=Z^2V_k+|U_{ei}(k)|^2\chi(k).
%         &=&Z^2V_k+|\Delta n_f(k)|^2/\chi(k,T).
\end{equation}
Its $r$-space form is obtained by Fourier transformation, and hence the
calculation of pseudopotentials and pair-potentials from the NPA
within LR is very simple, rapid and provides first-principles results.
The ion-ion LR pair potential $V_{ii}(r)$ will be denoted by
$V(r)$ for simplicity, where needed.
 
\begin{figure}[t]
\includegraphics[width=.95\columnwidth]{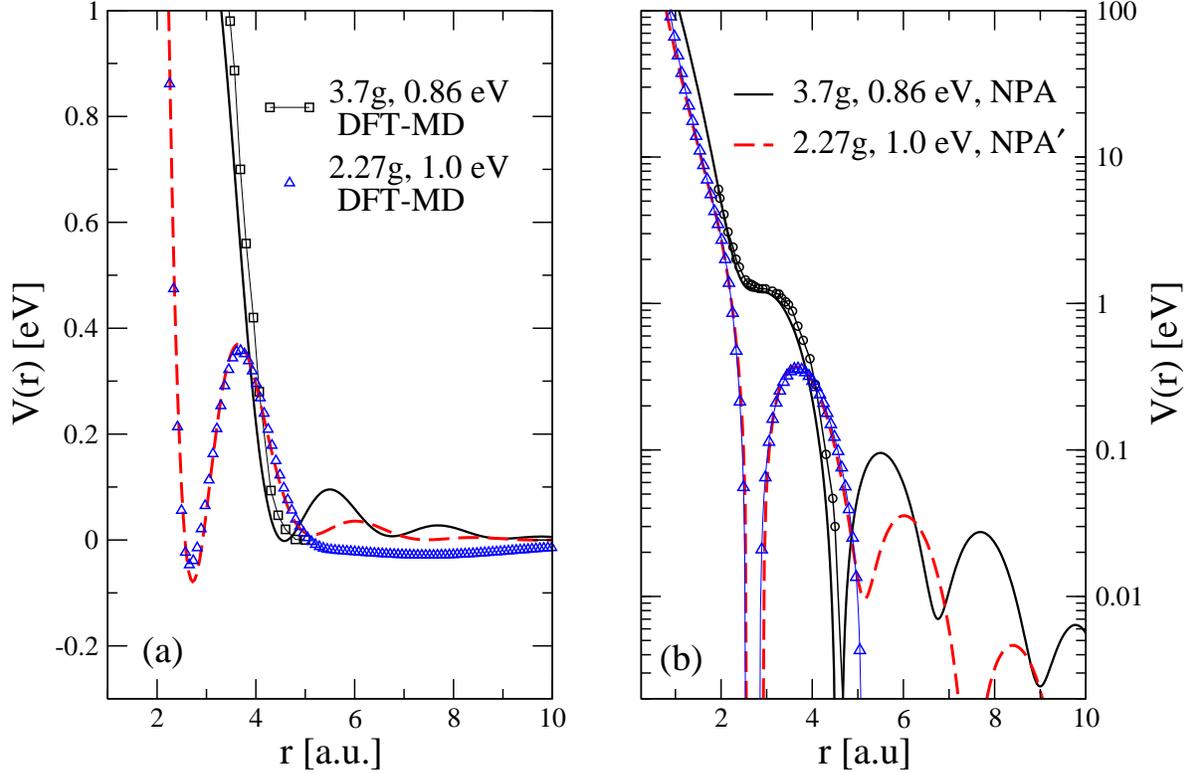}
\caption{ (Color online)  The C-C pair potentials from
a 64-atom force-matched QMD calculation and from the NPA.
(a) The QMD potential at
 $\bar{\rho}$=3.27 g/cm$^3$~\cite{whitley15}
compared with the NPA linear-response pair potential.
The QMD  potential from the $N$=64 simulation fails to recover the higher Friedel
oscillations (for $r>5$ a.u) seen in the NPA potentials and may not
capture long-range correlations relevant to phase transitions.
(b) The same data on a log-y scale. The maximum of the
$g(r)$ at this density corresponds to the
positive energy ledge for this range of densities.
\label{Vr.fig}
%Fig. 2
}
\end{figure}

The $N$-center QMD calculation
provides an $N$-center potential energy surface. This may be used to
construct force-matched pair-potentials~\cite{whitley15,Stanek21}. A
typical comparison is shown in Fig.~\ref{Vr.fig}.
The electrons have been
 eliminated from such potentials which require
three-center (and higher) terms in the total energy.
In contrast, the NPA
retains contributions from the two-component system of free electrons
and ions, pair-interactions and their xc-contributions to the
free energy.

 With $Z=4$,
$l$-carbon is a dense electronically degenerate metal even at 1 eV, even for
2.5 g/cm$^3$ when $E_F\simeq 23$ eV.
Hence electron-ion scattering occurs essentially at the Fermi
energy $E_F$, with a momentum transfer $q=2k_F$, where $k_F$ is
the Fermi momentum. The form-factor  $M_k$ changes sharply near
5 g/cm$^3$ as seen from Fig.~\ref{UeiSk.fig}(a).
Then, near
the `diamond density' $\rho_{\rm D}$, viz.,
$\simeq$ 3.52 g/cm$^3$
$M_k$ at $k=2k_F$ passes through zero and changes sign
(curve with circles).
So at $\rho_{\rm D}$, when $U_{ei}(k=2k_F)$ becomes zero
 the pair potential  becomes just $Z^2V_{k_F}$. Then the pair potential is
 that of two  {\it unscreened} point-like carbon ions, and
 interact strongly. Both NPA and QMD methods find
 a first-order LPT very near this density.

The NPA  structure factors $S(k)$ displayed in Fig.~\ref{UeiSk.fig}
 are in good agreement
with the QMD data, showing the split-peak structure etc., but become less
satisfactory at lower densities, i.e., 
below $\rho_{\rm D}$ and certainly below 3 g/cm$^3$, at 1 eV.  
Stanek  {\it et al}~\cite{Stanek21}  showed this explicitly
in their study of NPA pair potentials at the graphite density.
This is consistent with the fact that liner-response methods
are known to become unsatisfactory for expanded metals.

\section{Strong 2$k_F$ scattering and the Friedel structuring of the fluid}
The ion distribution $\rho(r)=\bar{\rho}g(r)$ is given
by the NPA-DFT equation (see Appendix) for ions  which are classical.
 It uses the pair-potential
 $V_{ii}(r)$ and an ion-ion xc-functional based on hyper-netted
chain (HNC) diagrams~\cite{DWP82}. This amounts
 to solving a HNC equation~\cite{DWP82} using
the NPA pair potential to yield the
$g(r)$ and the $S(k)$ of the ion subsystem.  
In Fig.~\ref{UeiSk.fig}(b)
we compare the results from the NPA, and QMD ~\cite{Vorberger20}
 where the PBE xc-functional ($T=0$) was used.
The first peak of $S(k)$ displays lower-$k$ (Lk)
and higher-$k$ (Hk) subpeaks.

The inter-particle interactions are dominated by electron-ion interactions
 causing
  strong electron scattering from
one edge of the Fermi surface to the opposite edge, with a momentum
 transfer of $q=2k_F$ as the fluid
is essentially degenerate since $T/E_F$ is small. In a solid
this could have spawned a charge-density wave with the
period 2$k_F$. The  fluid
 responds by creating the
high-$k$ peak Hk as close to 2$k_F$ as possible for densities in the range
10 g/cm$^3 \ge\bar{\rho}\ge 3$ g/cm$^3$, as shown in Fig.~\ref{UeiSk.fig}(b).

The electron xc-energy favours a high-density electron fluid even
if it requires positioning ions in a positive energy
ledge of the pair-potential (e.g., Fig. 2). %~\ref{Vr.fig}).
This is less advantageous at lower density and the peak Hk
at $2k_F$ grows at the expense of Lk, as seen in Fig.~\ref{UeiSk.fig}(b).
Scattering from atoms positioned in the Friedel  minima need to
 contribute coherently
to form the subpeak Hk in $S(k)$ at 2$k_F$. This coherence
links the coordination number with the liquid phases of the fluid
and their LPTs.

The ion distribution
modifies itself to minimize the energy of the system
in two ways: (i) its Hk peak locates itself as
closely as possible to  2$k_F$. (ii) At densities below $\rho_{\rm D}$,
the first atomic shell attempts to  continue to
retain a high coordination number $N_c$ at the expense of outer shells.   
Even when $\bar{\rho}$ is low, the {\it local density} of the
fluid  adjusts to bring Hk to 2k$_F$, lowering the  free energy
 via  strong scattering at $E_F$. That is, atoms may be drawn
 towards the central ion by decreasing the depths of secondary
 Friedel minima. This is evidenced by the nature of the
pair-potentials at different density regimes, as discussed
in Sec.~\ref{lpt-origin.sec}.

The split structure of the first peak of the structure factor
of liquid carbon, liquid silicon~\cite{cdwSi20}
and liquid germanium~\cite{Aers-CDW-Gibb86,DWP-Carb90}
arise from similar physics. Thus we may consider liquid carbon to be a
prototype of a class of liquids whose ionic structure   
has momentum ordering due to strong  2$k_F$ scattering
and reflects the 2$k_F$ periodicity of the Friedel oscillations in
 the pair-potentials.

The structure of momentum-correlated fluids like $l$-C, $l$-Si and $l$-Ge is
already tightly linked by the strong 2$k_F$ scattering at the Fermi energy,
and further adjustment to how the atoms are packed in the fluid by, say,
bridge-diagram corrections is hardly possible. Thus the structure factors
and derived properties like the pressure are quite insensitive to bridge
 corrections~\cite{Aers-CDW-Gibb86}. This is in contrast to `normal'
metallic fluids like $l$-Al where the bridge corrections are important
enough to modify the pressure and the compressibility.

Simulations using QMD do not use an ion-ion xc-functional,
but use a large number of atoms to build the ionic correlations through
explicit interactions. They reveal the existence of transient microscopic
``bonding'' which may be the real-space {\it dynamic} manifestation of
 the 2$k_F$ correlations in momentum space that seem to occur in this
 class of fluids with Friedel ordering, leading to a $2k_F$ subpeak in
$S(k)$.

\section{LPTs in liquid carbon}
 The LPTs  are evidenced by the  discontinuities in the
  pressure and the compressibility obtained from the NPA. We
first make a series of calculations on a finely spaced set of points
in density along an isotherm using the NPA, profiting from the
rapidity of NPA  calculations. The $k\to 0$
limit of $S(k)$ is the isothermal compressibility ratio $\kappa/\kappa_0$,
where $\kappa_0= 1/\bar{\rho}T$ is the ideal (noninteracting) compressibility.
Such a determination of $\kappa$ is independent of the pressure
calculation. That is, $\kappa$ is not
determined by taking a  density derivative of the pressure, and
constitutes an independent determination of LPTs including those
that are second order transitions. Unfortunately, such a direct and
microscopic determination of $S(k\to 0)$ is very costly in QMD as the
accessible smallest value of $k$ is limited by the inverse of the
linear dimension $L\propto N^{1/3}$  of the simulation box.

\begin{figure}[t]
\includegraphics[width=0.95\columnwidth]{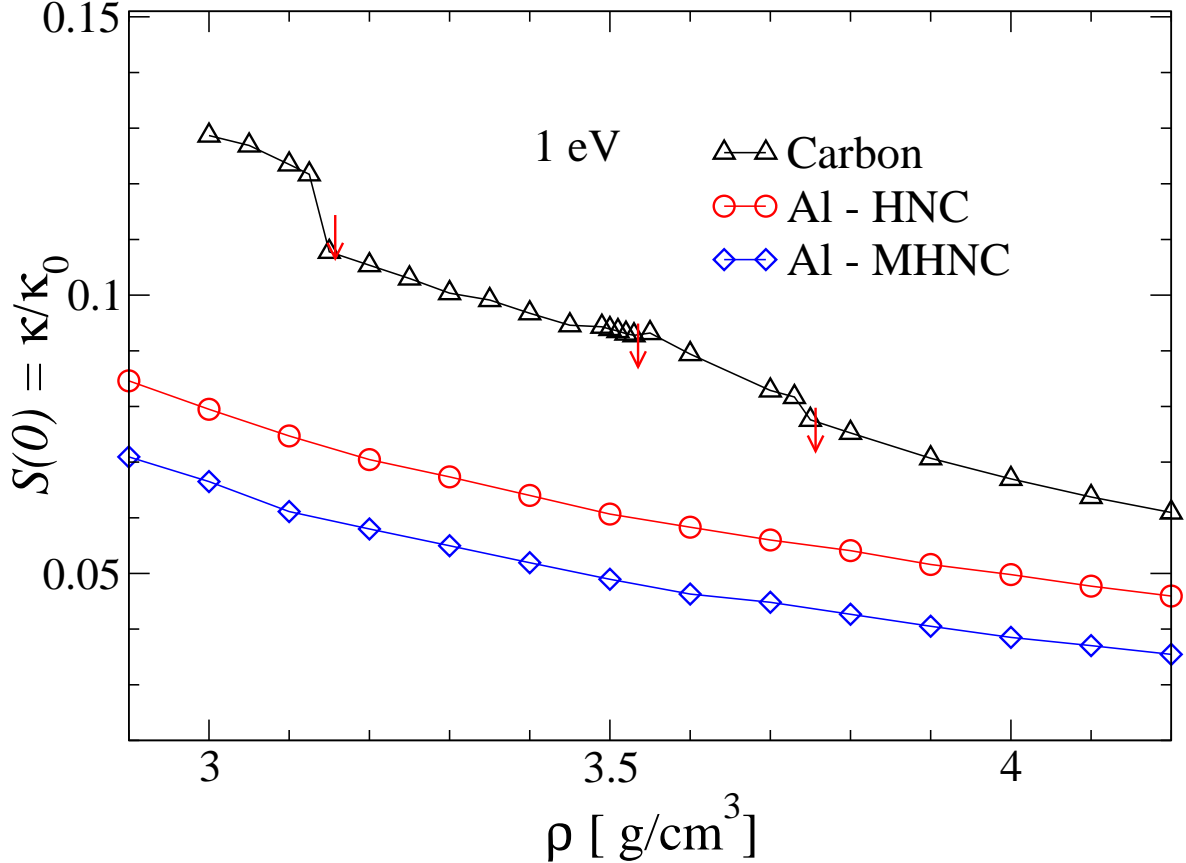}
\caption{
(Color online) The compressibility ratio $\kappa/\kappa_0$  of $l$-carbon
 and $l$-aluminum
 at 1 eV in the range of
 densities  3-4 g/cm$^3$ from NPA calculations. Their normal densities
are,  diamond $\sim 3.5$ g/cm$^3$, and aluminum, 2.7 g/cm$^3$.
The structure of  $l$-Al is not controlled by strongly correlated
 2$k_F$ scattering and  Friedel-oscillations, and  shows no
 discontinuities associated with changes in coordination number, in
contrast to  $l$-carbon displaying
such discontinuities (short vertical arrows). 
Bridge corrections modify the $S(0)$ of $l$-aluminum, as shown
by the difference in the HNC (without bridge) and MHNC (with bridge)
 calculations.  The $S(0)$ of
  $l$-carbon remains unchanged for this $\rho, T$ region
on inclusion of bridge corrections (see Appendix).
\label{S0-C-Al.fig}
}
\end{figure}

In Fig.~\ref{S0-C-Al.fig}.
We show the compressibility ratio for $l$-carbon as well as
for $l$-aluminum for the range  $\sim$ 3-4 g/cm$^3$.
  Aluminum forms a metallic liquid
with $\bar{Z}=3$  but its structure factor does not show the 2$k_F$
splitting seen in $l$-carbon or $l$-silicon, and hence there is
no incipient charge-density-wave type of effect. The Friedel
ordering which links the coordination and shell filling does not
occur in liquid aluminum where the liquid structure is mainly
determined by packing effects. This also makes Al sensitive to
bridge corrections.Thus HNC calculations of the $S(0)$ for $l$-aluminum
differs from the result obtained from the modified HNC
equation(MHNC) or on using MD. In contrast, the highly Friedel-correlated
$l$-carbon is insensitive to bridge corrections as atomic positions
preferably pack into the shells formed by the minima of the
 Friedel  oscillations in the pair-interactions. Unlike
$l$-aluminum, $l$-carbon shows three discontinuities in this
range of  densities. We study them in detail below.

\begin{figure}[t]
\includegraphics[width=0.95\columnwidth]{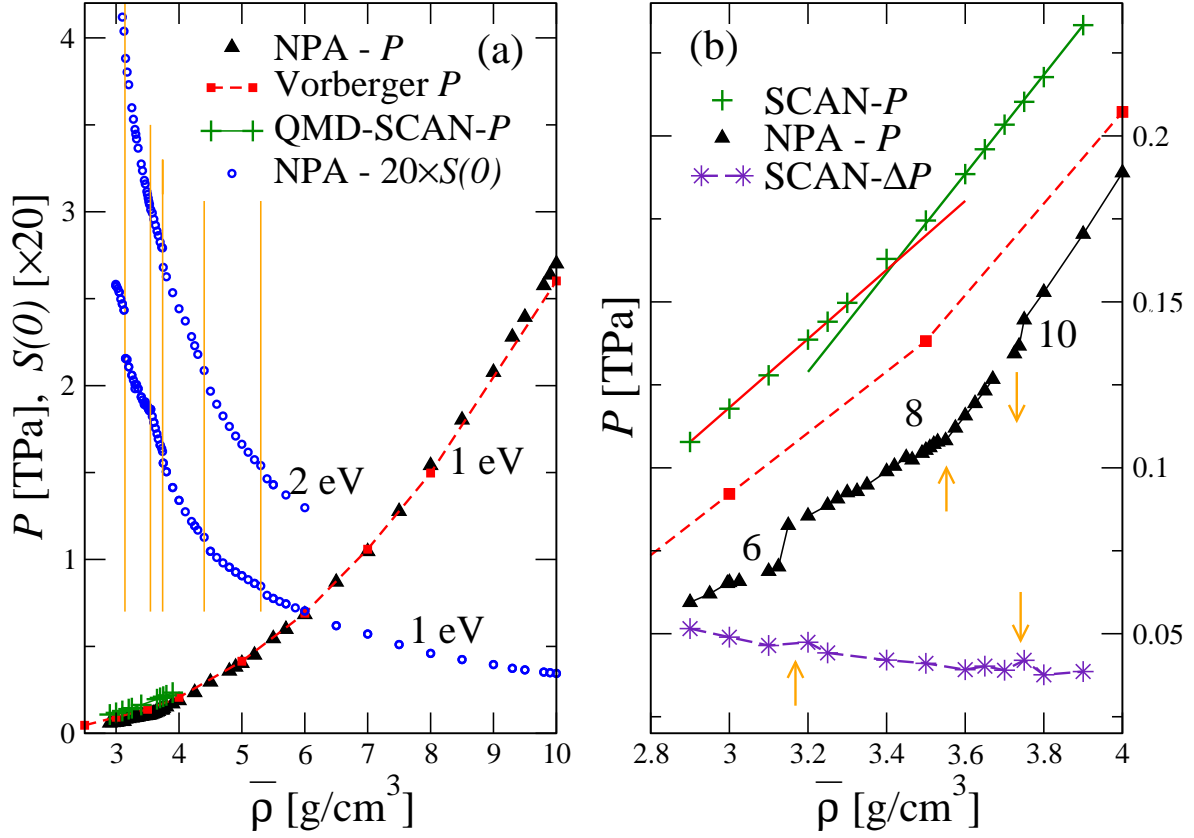}
\caption{
( Color online)  Evidence for LPTs in $l$-carbon
from QMD and NPA calculations.
(a) The pressure at 1 eV
 from QMD-PBE~\cite{Vorberger20},
QMD-SCAN and NPA calculations.
 The $k\to 0$ of $S(k)$, viz., $S(0)=\kappa/\kappa_0$ where
$\kappa$ is the isothermal compressibility. It
displays discontinuities (vertical lines) at the LPTs.  
(b) The discontinuities in the range 2.9 g/cm$^3<\bar{\rho}<4$ g/cm$^3$
from the NPA pressure indicate
 LPTs at $\bar{\rho}\sim$  3.15 g/cm$^3$ (LPT$_3$),  3.52g/cm$^3$ (LPT$_{3.5}$), 3.75 g/cm$^3$ (LPT$_{3.7}$.
The data points above and below $\bar{\rho}\simeq$ 3.42 g/cm$^3$
 from 108-atom QMD-SCAN simulations fall accurately on straight
 lines intersecting
 at $\bar{\rho}\simeq$ 3.42 g/cm$^3$ if a single Gaussian is fitted
to the QMD $P$-distributions. However, the width of the
 SCAN-QMD $P$ distributions (labeled SCAN-$\Delta P\times$3.3)
increases  (arrows) at LPT$_3$ and LPT$_{3.7}$ and are best
fitted with two-Gaussians. See Fig.~\ref{pGaus.fig} for two
densities near LPT$_3$.
\label{pS0.fig}
%fig. 3
}
\end{figure}

In Fig.~\ref{pS0.fig}(a) we display the compressibility ratio $S(0)$
and the pressure at 1 eV for a larger range of densities. The
 compressibility ratio at 2 eV is
also shown. The discontinuities are indicated by
vertical lines. A selection of the discontinuities are studied
 further using the vastly  more expensive  QMD simulations implemented
via VASP.
The SCAN electron xc-functional at $T=0$ was used and the results
are presented in Fig.~\ref{pS0.fig}(b).
 The use of the $T=0$
approximation in QMD is quite accurate as $T/E_F$ is
 0.035 even at $\bar{\rho}$
= 3 g/cm$^3$ at 1 eV. We identify the following discontinuities and
label them as follows.

The LPTs are named  LPT$_3$, LPT$_{3.5}$, LPT$_{3.7}$, LPT$_{4.5}$
  and LPT$_{5.3}$
being at  densities 3.15, 3.5, 3.75, 4.5 and 5.3 g/cm$^3$ respectively
as given by the NPA calculation.
They can be conveniently classified in terms of the
coordination number $N_c$ (as estimated from the area under the
 first peak of the  pair-distribution functions) of the regions separated
by the LPTs. Beyond the LPT$_{5.3}$ $N_c$ has
reached 12 and no further LPTs associated with changes in coordination
number are expected.  Vorberger {\it et al}~\cite{Vorberger20} have
 reported estimates of the coordination number in the relevant density range,
 and our results are consistent with their $N_c$ estimates. They are
indicated in Fig.~\ref{pS0.fig}(b) where we have selected the density range
2.9 g/cm$^3<\bar{\rho}<4$ g/cm$^3$ for further detailed study.

Figure~\ref{pS0.fig}(a) displays the QMD pressures  
from Ref.~\cite{Vorberger20}, and from our QMD runs
 using the SCAN functional and 108 atoms. The NPA uses the finite-$T$
 xc-functional of Ref.~\cite{PDWXC} within the local density
approximation (LDA). We have compared the finite-$T$
electron xc-functional that we use in the NPA  with that of
 Dornheim {\it et al}~\cite{Dornheim18,cdw-N-rep19},
and find excellent agreement in the range of $\bar{\rho},T$
used here. The LDA is deemed to be adequate because the
charge density in the NPA is very simple compared to the
complex $N$-center density used in QMD calculations.
The use of ever more complex $N$-center calculations
have led to a ``Jacob's ladder'' of ever-more complicated
electron xc-functionals that are deployed in QMD.
 
 The pressure estimates from
PBE-QMD, SCAN-QMD and the LDA-NPA calculations fall within the
differences in the xc-functionals used. The NPA calculates the
pressure from a  free energy calculation of the interacting system.
Faussurier {\it et al}~\cite{Fauss21} have noted that the VASP
code  estimates the ionic contribution to the pressure using an ideal-gas
approximation and a stress-tensor evaluation, rather than from the total
free energy. This may also contribute to a difference between the
 QMD and NPA estimates of the pressure (where no ideal gas
 approximation is made).
Furthermore, the NPA recovered the Ganesh-Widom pressure for
$l$-silicon close to its melt line accurately~\cite{cdwSi20}.
The Ganesh-Widom calculation implemented its own
free energy calculation to obtain the pressure. So, the
full power and rigour of QMD calculations may, or may not
be available depending on how QMD codes implement the finite-$T$
pressure calculations.

The discontinuities in $P$, and in $S(0)$ independently support one another.
The behaviour of S(0) at $T$= 2 eV is also indicated in Fig.~\ref{pS0.fig}.
The 1 eV pressure isotherm obtained from the QMD-scan calculation as
 well as that from the NPA are shown. The QMD results (obtained using the
 SCAN functional) independently
  confirm the LPTs found using the NPA,
 but at slightly shifted
 densities. Thus the NPA-LPT$_{3.5}$ occurs at 3.42 g/cm$^3$ in the QMD-SCAN
 data; this LPT corresponds to a structural adjustment
when the form factor $M_k$ changes sign (Fig.~\ref{UeiSk.fig}),
while $N_c$ remains at 8. The
LPT$_{4.5}$ is also a structural adjustment while $N_c$ remains at 10.
The origin of these LPTs is discussed in Sec.~\ref{lpt-origin.sec}.

 The QMD-SCAN pressure has a sharp break near the LPT$_{3.5}$.  
However,  the other two LPTs seen in the NPA are not seen in the VASP
pressure  calculation.  Whether this arises from unknown  approximations in the
NPA approach, or in the
VASP calculation of the ionic contribution to $P$ ~\cite{Fauss21}, or
from the inadequate treatment of long-ranged effects (c.f., fig.~\ref{Vr.fig})
due to the finite-size of the simulation ($N$ = 108) in QMD,  is not clear.

 However, a deeper  examination reveals
evidence of these two LPTs even in the QMD data.
The pressure {\it fluctuations} $\Delta P$ are expected to
form a Gaussian distribution (see Appendix) for a uniform equilibriated fluid
(Fig.~\ref{pGaus.fig}).
 The width $\Delta P$ in the QMD-SCAN (labeled SCAN-$\Delta P$)
changes significantly
 near the  discontinuities (vertical arrows, Fig.~\ref{pS0.fig}(b))
 at
LPT$_3$ and LPT$_{3.7}$. They are best fitted to a sum of two Gaussians,
(as seen in Fig.~\ref{pGaus.fig}) unlike at LPT$_{3.5}$.
 The need for two-Gaussians at LPT$_3$ and
LPT$_{3.7}$ to describe the QMD pressure fluctuations provides some
confirmation of the LPTs
explicitly seen in the NPA calculation. However, phase transformations involve
the cooperative action of large numbers of atoms, and hence we have
also examined the $N$-dependence of the pressure estimates
from the QMD simulations in the regime of densities where LPTs are found.
This is discussed in the appendix and confirms that the $N$=108 QMD
simulations probably provide converged values of the pressure.

\begin{figure}[t]
\includegraphics[width=.94\columnwidth]{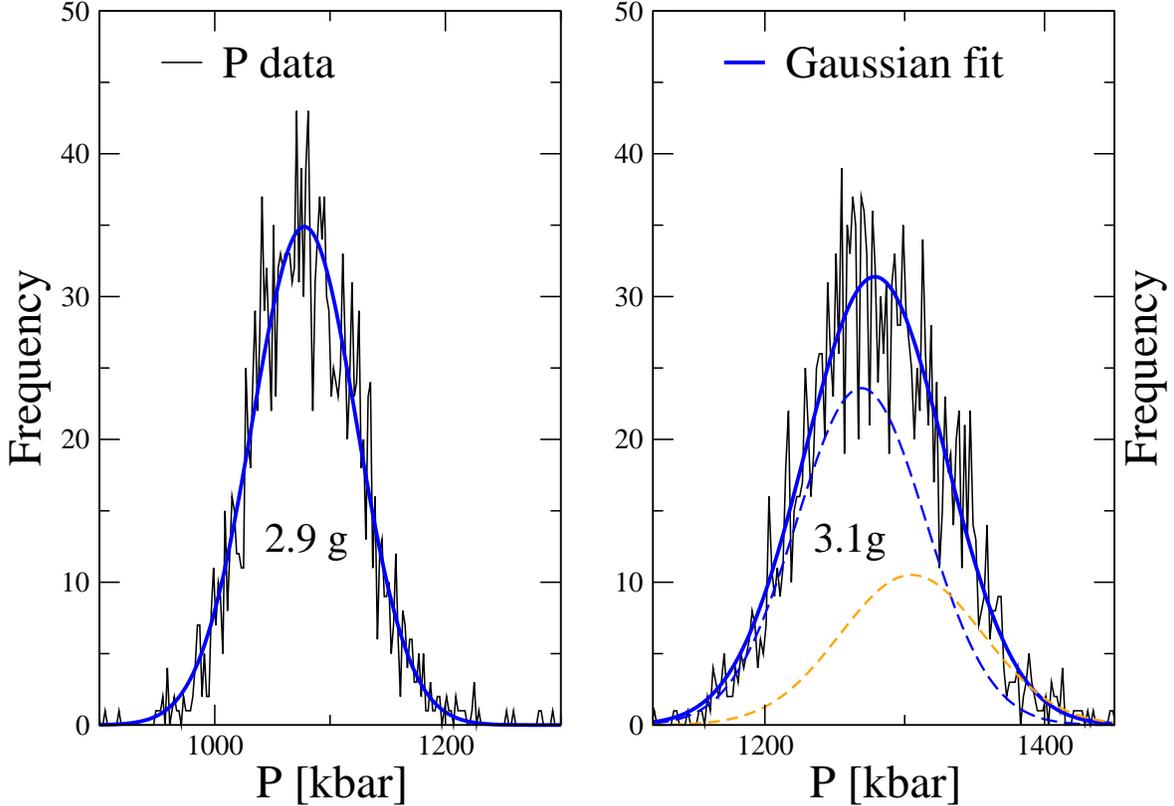}
\caption{
(Colour online)  The $P$-distribution from $N=108$ atom QMD-SCAN calculations.
(a) 2.9 g/cm$^3$ and (B) 3.1 g/cm$^3$, at LPT$_3$. While the 2.9 g/cm$^3$
case fits to one Gaussian, that at 3.1 g/cm$^3$  best
fits to two Gaussians, implying an inhomogeneous system at the
LPT in the 108-atom QMD simulation. The LPT$_{3.7}$ is also bimodal, but the
LPT$_{3.5}$ is not  bimodal (for more
details, as well as results from $N=64$ simulations, see the Appendix).
 \label{pGaus.fig}
% Fig.~4
}
\end{figure}

\section{The origin of discontinuities in the NPA pressure}
\label{lpt-origin.sec}
Many average-atom (AA)
models~\cite{Anta00,Murillo13,Rosznyai2008,StarretHam14,YongHou-AA-17}, also
sometimes referred to as ``atoms-in-jellium'' models, confine
 all the bound and free electrons of an ion inside the Wigner
 Seitz sphere. This introduces an electron chemical potential different from
the non-interacting chemical potential $\mu_0$ required
 in  density functional models.
In DFT the interacting electrons are mapped onto a system of non-interacting
electrons (i.e., $\mu_0$ is applicable) at the {\it interacting} density.
Thus AA  models that use a $\mu\ne\mu_0$ are not strictly DFT  models.

 In the NPA the electrons are not
confined to the WS-sphere, but to the correlation sphere which has a
radius large enough to ensure that all pair-distribution functions
 have reached unity and all correlations have died out. Typically
$R_c\sim 10r_{\rm ws}$ to $5r_{\rm ws}$ and $\mu_0$ is the correct
chemical potential for the NPA, as required by DFT. Furthermore,
simple AA models have
 to deal with
electrons that ``leak out'' of the WS-sphere, and may produce
model dependent effects
arising from the choice of boundary conditions at the
surface of the WS-sphere; these do not arise in the NPA.

However, NPA models as well as AA models have to deal with discontinuities
that arise  when a well-confined
bound state moves upwards in energy and into the continuum due to changes in
 density or temperature. In some cases, such ionization is
accompanied by phase transitions and discontinuities in physical
properties~\cite{CPP-carb18}. In other cases, when unphysical
 discontinuities occur, they can be
corrected by ensuring that appropriate discontinuities that cancel them
are properly included in the  xc-functionals. However, there are no
changes in the degree of ionization ($\bar{Z}$) or any other critical
parameters for $l$-carbon  in the range of densities studied here, and hence no
spurious discontinuities are expected in this range of $\bar{\rho},T$.

\begin{figure}[t]
\includegraphics[width=0.9\linewidth,keepaspectratio]{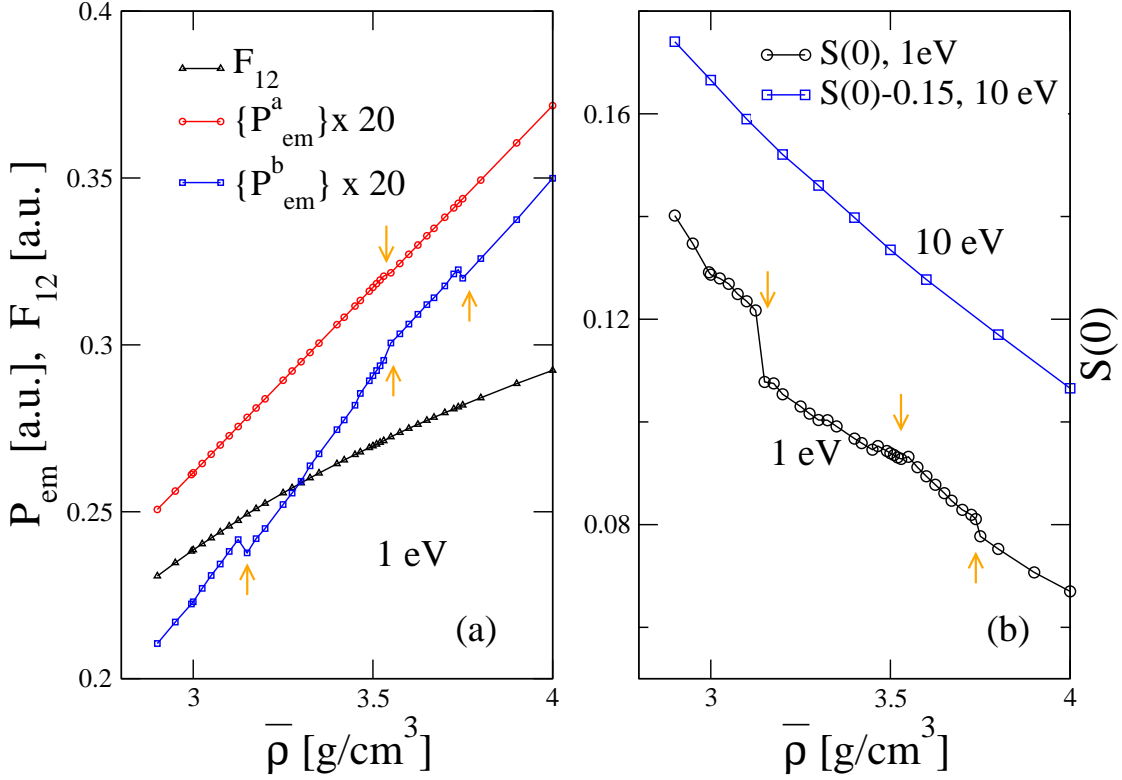}
\caption{
(Color online) (a) The two components of the
embedding pressure obtained (without numerical differentiation) from the
 embedding
free energy of the neutral pseudo atom. The pair-interaction
free energy $F_{12}$  containing bonding effects is also displayed.
 The embedding pressure component $P^b_{em}$
displays three discontinuities, while  $P^a_{em}$ shows a slight
 discontinuity near
$\bar{\rho}=3.52$. No discernible discontinuity is seen in the bonding
 free energy
 $F_{12}$, showing that the LPTs are not an effect associated with
bonding or pair interactions.
(b) The $k\to 0$ limit of the structure factor at 5 eV, 3 eV and 1 eV.
 $S(0)=\kappa/\kappa_0$,where $\kappa_0, \kappa$ are the ideal fluid
 compressibility and the interacting compressibility respectively.
 The value of $S(0)$ at 5 eV
is shifted  by 0.09 and $S(0)$ at 3 eV by 0.08.
$S(0)$ at 2 eV was displayed in Fig.~\ref{pS0.fig}.
\label{Pem.fig}
}
\end{figure}

The NPA pressure is obtained from an evaluation of the Helmholtz free
energy $F$ of the system.
This consists of contributions of the form~\cite{Pe-Be,eos95}:
\begin{equation}
F=F^0_e+F^{xc}_e+F^a_{em}+F^b_{em}+F_{12}+F^0_I.
\end{equation}
The first two terms deal with the free energy of the non-interacting
uniform electron fluid and its finite-$T$ exchange-correlation energy
 at the given
 density and temperature $T$. The last term, $F^0_I$ is the
ideal (classical) free energy of the ion subsystem. The third and fourth
 terms
 together
form the embedding energy of the nucleus and the inhomogeneous electron
density that form the neutral pseudo-atom in the uniform electron fluid. The
fifth term, $F_{12}$ contains the interactions between pseudoatoms brought
in via the pair-potential, pair-distribution functions and the
ion-ion correlation effects. The contribution to the pressure from all
the terms except $F_{12}$ can be expressed analytically. The density
derivative of the term $F_{12}$ has to be evaluated numerically to give a
$P_{12}$. Then the pressure is:
\begin{equation}
P=P^0_e+P^{ex}_e+P^a_{em}+P^b_{em}+P_{12}+P^0_I.
\end{equation}
No discontinuities are expected in $P^0_e, P^{ex}$ or $P^0_I$ when treated
as a function of $\bar{\rho}$ as the number of free electrons per ion,
 $\bar{Z}$, remains constant at a value of four in this density range. Hence,
in Fig.~\ref{Pem.fig}, panel (a) we display only the two components of the
embedding pressure and the pair free energy $F_{12}$ as a function
of $\bar{\rho}$ at 1 eV.

In order to understand the physical content of the
embedding pressure (see Ref.~\cite{Pe-Be}) that contain the discontinuities,
we define
the symbol $\star$ to define a convolution
product. We also define the integration over all space via the symbol $\circ$.
 That is:
\begin{eqnarray}
f\star g&=&\int f(\vec{r})g(\vec{r}-\vec{s})d\vec{s}\\
f\circ g&=&\int f(r)g(r) d\vec{r}.
\end{eqnarray}
The volume of integration is a sphere of radius $R_c\sim 10r_{\rm ws}$ at
low $T$ and $\sim 5r_{\rm ws}$ at higher $T$.

Then the potential
that produces a single neutral pseudoatom in the uniform
fluid of mean electron density $\bar{n}$ and mean ion density $\bar{\rho}$,
is:
\begin{equation}
\label{v-npa.eqn}
V^{npa}=\frac{1}{r}\star(-Z\delta_0+\nu+\Delta n).
\end{equation}
Here $Z\delta_0$ defines the nuclear term at the origin,
while $\nu$ (i.e.,  $\nu(r)$)
 is  the density of the spherical cavity which mimics the ion
 distribution $\rho(r)$ by
 $\bar{Z}g_{cav}(r)$.
\begin{equation}
\nu(r)=\bar{n},\, r>r_{\rm ws}, \mbox{ else } \nu(r)=0.
\end{equation}
Also,  $\Delta n$ is the displaced electron density with reference
to the mean electron density $\bar{n}$. Then, denoting the volume of the
 Wigner-Seitz sphere by $\Omega_{\rm ws}$,
the two embedding pressures can be written as \cite{Pe-Be}:
\begin{eqnarray}
P^a_{em}&=&(\nu-\bar{n})\circ V^{npa}/\Omega_{\rm ws}\\
P^b_{em}&=&-\bar{Z}V^{npa}(r_{\rm ws}).
\end{eqnarray}
Since $\bar{Z}=4$, and remains at that value through out the range of
densities studied, the discontinuities in the embedding pressure are
caused solely by the `external potential' associated with the creation
of the carbon pseudoatom and its pseudopotential
 (c.f., Fig.~\ref{UeiSk.fig}).

This re-enforces our conclusion that the observed LPTs
are associated with changes in coordination number $N_c$ and the effect on
the electron-ion interactions at the Fermi energy via $q=2k_F$ scattering.
The coordination number $N_c$ can be approximately determined from the
area under the first peak of $g(r)$, and has been reported in detail
by Vorberger {\it et al}~\cite{Vorberger20}. Changes in $N_c$ do  not
seem to appear as
discontinuities is QMD $P$ but clearly cause discontinuities in
the NPA pressure and the NPA compressibility i.e., a derivative of
the pressure which is independently calculated via $S(0)$. The long
range of the NPA potentials seems to play a significant role here.\\

\begin{figure}[t]
\includegraphics[width=.94\columnwidth]{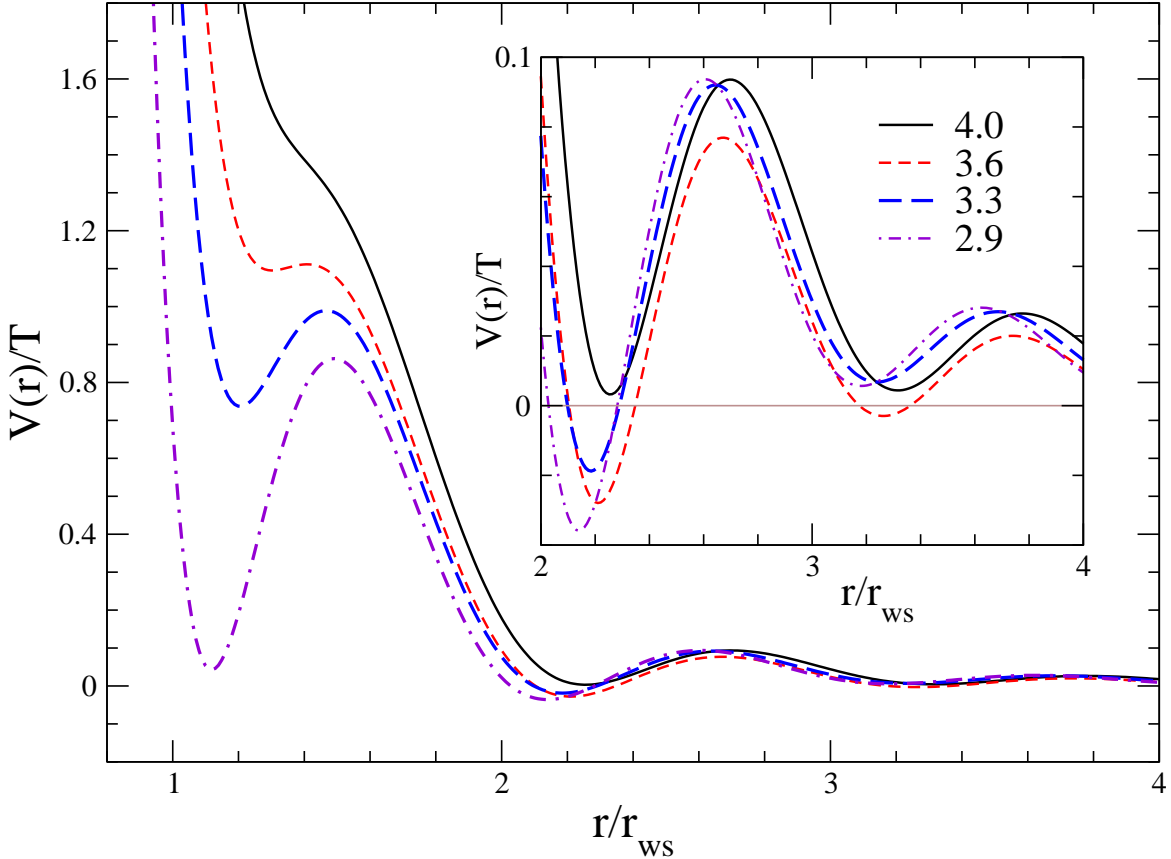}
\caption{
(Color online)  C-C pair potentials at densities in
regions separated by the discontinuities in the
 pressure. The densities shown ($4,\, 3.6,\, 3.3,\, 2.9$ g/cm$^3$)
are for regions where the coordination numbers $N_c\sim 10, 8, 6,$ and
$N_C<6$ prevail. The discontinuities imply LPTs
in the NPA calculation. The QMD data show a discontinuity near
the the diamond density, but only broadenings in the pressure fluctuations
at the other LPTs, as displayed in Fig.~\ref{pGaus.fig}.
\label{potsLPT.fig}
}
\end{figure}

The role of the long-ranged Friedel oscillations in ``ordering the fluid''
is made clearer by an examination of the potentials in the density ranges below
$\rho_{\rm D}$, near $\rho_{\rm D}=3.5$ g/cm$^3$ and above $\rho_{\rm D}$.
These are provided in Fig.~\ref{potsLPT.fig} which is  discussed
below.

\section{Discussion}
The structure factor  data show that the
structure of $l$-carbon is determined by strong
electron-ion scattering across the Fermi surface, and by the
Friedel oscillations of the pair-potentials acting in consort to
produce a subsidiary peak in $S(k)$ at twice the Fermi wavevector.
 This `Coulomb fluid' model
of $l$-carbon provides a complete account of the structural and thermodynamic
properties of $l$-carbon and its liquid-liquid
phase transitions  without invoking covalent bonds, using only
simple $s$-wave  pair-potentials.
The covalent bonds are mere transients, lasting
the lifetime of longitudinal phonons, while DFT deals with
thermodynamic ($\omega\to 0$) averages.

The LPT$_{3.5}$ near $\bar{\rho}_{\rm D}$ is evident in
 the QMD-SCAN $P$ data that accurately fall on two straight lines
crossing at 3.42 g/cm$^3$. The LPT
occurs almost at the nominal
`diamond' density, when $U_{ei}(2k_F)$ becomes zero thereby creating
a strong ion-ion interaction.
A discontinuity in the electrical conductivity is also found
at this density (see the last figure in the Appendix).

The minima in the pair potential $V(r)$ (see Fig.~\ref{potsLPT.fig})
are all positive  in the high density region (e.g., $\bar{\rho}=4$ g/cm$^3$),
 with $N_c\sim$ 10 or more. The next lower density region develops a deeper
 first minimum and  negative secondary minima. The third region (e.g.,
 $\bar{\rho} =3.3 $g/cm$^3$) pulls in ions  to the center by making the
 secondary minima less attractive. In the lowest density region
the first minimum becomes  deep, and eventually becomes negative enough
 (e.g, exceeds twice the thermal energy) to form persistent covalent bonding.

The NPA calculation is for a uniform fluid
where the 2$k_F$ scattering is tightly linked to the liquid structure
and the coordination number through the Friedel  minima  in
the pair potential. They correlate the whole fluid as in a charge-density
wave that could have existed in a solid. It is this tight coupling of the
ionic structure to the Friedel oscillations that lead to LPTs at each
change of coordination number, a phenomenon common to $l$-Si~\cite{cdwSi20}.
Previous studies of LPTs~\cite{glosli99,WuLPT02} using QMD, by the
very nature of its  methodology  emphasized real-space
short-ranged  chemical bonding (e.g., $sp^3\to sp^2$) with short life
times and did not address the physics of 2$k_F$ scattering and related
`Fermi-liquid' phenomena.

{\bf In conclusion}, the ionic and electronic structure of $l$-carbon can be
accurately and inexpensively modeled as that of a uniform {\it liquid metal}
 whose
structure is dominated by strong electron-ion interactions at the Fermi surface.
We find three liquid-liquid phase transition at 1 eV in the range 2.9-4.0 g/cm$^3$
confirmed by both NPA and QMD calculations, and two higher-density LPTs.
 These results and similar results for $l$-Si~\cite{cdwSi20} suggest that
 metallic fluids
 with a peak in $S(k)$ at 2$k_F$ are
strongly Friedel correlated liquids that display multiple LPTs.

Our conclusion
that $l$-carbon is a `good metal' is of interest to long-standing controversies
 regarding the magnetism of giant planets~\cite{Ross81}
and also white dwarfs~\cite{Dufour07}.
Hence these results for carbon are relevant to terrestrial as well as
astrophysical systems, and critical to  emerging
 carbon materials~\cite{Hull20}.
\newline

{\bf DATA AVAILABILITY}

All the data used in this paper are available within the article in graphical
 form in the figures 1 to 10. If there is any difficulty in extracting them from
 the figures,  the data can be provided on request from the authors.

\appendix*

\section{}
$\,$\\
This Appendix  addresses the following topics:\\
\begin{itemize}

\item Details of the two density functional
methods (namely, QMD and NPA methods) used here and results
for $N=108$ and $N$=64 simulations.
\item  The NPA pseudopotentials and pair-potentials.
\item Non-subjective estimation of the pressure from QMD runs and evidence to
support the existence of a more broadened type of pressure distribution at
liquid-liquid phase transitions LPT$_3$ and LPT$_{3.7}$.\\
\item  The electrical conductivity of liquid carbon.
\end{itemize}

\subsection{Details of the two density functional
methods}
The theoretical methods used in this study
are (a) one-atom DFT as implemented in the NPA,  and (b) many-atom DFT (QMD) simulations
 using $N$=108 and $N$=64 carbon ions and  the associated number
of electrons assuming two bound electrons per ion.
The NPA is an `all-electron' method
while the QMD used here employed a pseudopotential and hence the 108 ion
simulation included 432 electrons, while the $N=64$ included 256 electrons.

\subsection{\bf The QMD simulations}
These  are finite-$T$ DFT-MD (QMD) calculations
where classical molecular dynamics is used to evolve 108 ions in a cubic
simulation cell, while the electrons, with four ionized
electrons per carbon atom  are treated quantum mechanically using
density functional theory.
The  numerical code implemented  in the
 Vienna ab initio simulation
package VASP 5.4.4~\cite{VASP}, and the projector augmented-
wave pseudopotential for the interaction between the nuclei
and the electrons provided in VASP were used. The exchange and correlation
 potential
 is approximated by the SCAN functional which has been found to
perform better for systems with covalent
 interactions~\cite{Remsing17}. The simulations used
 an energy cutoff of 414 eV  and a simulation time of 4 ps.
The ion temperature was controlled with a Nos\'{e} thermostat,
and  enough empty bands were included to ensure that the highest
energy bands had negligible occupations. Evaluations of the
Brillouin zone were performed at the Baldereschi
 $k$-point~\cite{Baldereshi73}.\\

The pressure obtained by the $N=64$ and 108 calculations are
given in Table~\ref{qmd-p.tab} for typical densities used
 in the main study.
\begin{table*}
\caption{\label{qmd-p.tab} The pressure in kB at 1 eV obtained from
QMD calculations using $N=64$ and $N=108$ carbon ions in the simulation.}
\begin{ruledtabular}
\begin{tabular}{cccccccccc}
$\bar{\rho}$, \;  g/cm$^3$ &3.0&3.1&3.2&3.3& 3.6&3.7&3.8&3.9\\
$P,\;N=64$, \;  kB& 1182& 1277& 1383& 1500& 1887& 2030& 2182& 2338\\
$P,\;N=108$, \;  kB& 1178& 1279& 1386& 1506& 1885& 2034& 2177& 2334\\
\end{tabular}
\end{ruledtabular}
\end{table*}
These pressure values suggest that the $N=108$ simulation is adequately
 converged for our purposes. The pressure at the lower densities fall
onto a straight line, while those at the higher densities fall on a
different straight line, with the pressure discontinuity occurring at
the critical concentration $\rho_{cr}$=
3.40 g/cm$^3$ for $N=64$, while the $N=108$ data place the discontinuity
at 3.42 g/cm$^3$ (see Fig.~\ref{pS0.fig}). The simulation box size
varies as $N^{1/3}$. Extrapolation of the two values
of $\rho_{cr}$ as a function of $1/N^{1/3}$ gives the large-$N$ limit to be
3.524g/cm$^3$, consistent with the NPA estimate of $\rho_{cr}$=3.52 g/cm$^3$.    

\subsection{\bf The NPA calculations} 
For a fluid, these are essentially like a self-consistent
atomic structure calculation for a single nucleus placed
in averaged-out spherically symmetric electronic and ionic densities, and have
strong similarities to an average-atom
 calculation~\cite{Anta00,Murillo13,Rosznyai2008,StarretHam14,YongHou-AA-17}.
The environment is described by one-body densities
$n(r)$ of electrons, and $\rho(r)$ of ions,
rather than by explicitly enumerated atoms (as used in QMD). The
one-body densities $n(r)$  and $\rho(r)$ are
self-consistently evaluated using Kohn-Sham type finite-$T$
DFT equations. In addition to the quantum Kohn-Sham
equation derived from the stationary condition
on the grand potential $\Omega([n],[\rho])$ for functional derivation
with respect to the electron density $n(r)$, we
use the stationary condition for variations of the ion density
as well. Thus,
\begin{equation}
\delta \Omega(n,\rho)/\delta \rho = 0
\end{equation}
is used in the theory of the NPA method. This involves an ion-ion xc-functional
where the exchange part is zero since the ions can be treated as
classical particles in most cases. Furthermore, instead of the Kohn-Sham equation,
we get a classical DFT equation~\cite{Chihara78,EvansDFT79} that
 can be identified as a
Boltzmann-like distribution for $\rho(r)$ that has to be solved
simultaneously with the Kohn-Sham equation~\cite{DWP82} for the
electron density $n(r)$. This can be done
either using MD, or using an integral equation approach exploiting
the spherical symmetry of the fluid.
\begin{equation}
\rho(r)=\bar{\rho}\exp\{-V_{\rm cKS}(r)\}.
\end{equation}
The classical Kohn-Sham potential $V_{\rm cKS}(r)$ occurring in the
above equation can be identified with the potential of
mean force used in the theory of liquids. Then the ion-ion
xc-potential can be seen to equal to the sum of hyper-netted-chain
diagrams and bridge diagrams, if the electron-ion  xc-potential
is neglected~\cite{DWP82}.
The electron-ion
xc-functional $F^{xc}_{ei}$ is usually neglected in most NPA
calculations, being largely equivalent
to making the Born-Oppenheimer approximation, and neglecting
certain correlation corrections  of the
from $\langle n({\bf r})\rho({\bf r}')\rangle-
\langle n(r)\rangle\langle\rho(r')\rangle$. This is
equivalent to using a `random-phase type' approximation for the
electron-ion response function in regimes where we have
already used the linear-response form for the electron
response. This neglect of $F^{xc}_{ei}$ is quite appropriate for
dense  uniform fluids of carbon studied here.

The bridge contributions can be included using the hard-sphere model, with the
hard sphere packing fraction parameter $\eta$ selected using the Lado-Foils-Ashcroft
criterion~\cite{LFA83}. We find that the inclusion of bridge corrections
has only a negligible effect for the calculation
of the pressure or the compressibility of  $l$-carbon in the range of $\bar{\rho},T$
studied here, unlike for $l$-aluminum.
Typical values are given below.
\begin{table}
\caption{\label{pressure-bri.tab}. The NPA pressure at 1 eV without (labeled no B)
 and with (labeled with B)
bridge corrections, and the corresponding
hard-sphere packing fraction $\eta$ for selected densities.}
\begin{ruledtabular}
\begin{tabular}{lccc}
$\rho$ g/cm$^3$   & 3.0  & 4.0  & 8.0 \\
\hline \\
$P$ TPa, no B    &0.03413 & 0.1705 & 1.540\\
$P$ TPa, with B  &0.03359 & 0.1700 & 1.538\\
$\eta$           &0.1080  & 0.2030 & 0.3515\\  
\end{tabular}
\end{ruledtabular}
\end{table}
For any given case, the
whole calculation including the evaluation of pair-potentials,
structure factors, pressure, compressibilities, electrical and
thermal conductivities
can be carried out on a small laptop within a few minutes
of computational time. The small-$k$
limit of $S(k)$ is directly available from these calculations,
unlike in QMD calculations where the finite-size $L$ of the
simulation box limits the smallest available $k$ value to
$\sim \pi/L$. An HP-folio9470m laptop
has been used for NPA computations. Calculations for over 60 densities each were
carried out on isotherms of  1 eV, 2 eV, etc., while somewhat fewer densities
were studied at 0.8 eV, 3 eV, 5 eV and 10 eV. A sample of those
results are presented here.

Detailed discussions of the NPA may be found in several recent
publications~\cite{Stanek21,cdwSi20}, as well as in some of the earlier
 publications~\cite{DWP82,Pe-Be,eos95}. In this study, the density $\bar{\rho}$,
and temperature $T$ (in energy units) are such that the carbon atom
carries only the 1$s$ shell of bound electrons, providing a very simple model of an
atom in its electronic and ionic environment, and the corresponding
NPA equations are discussed in detail in Ref.~\cite{cdw-Carbon10E6-21}.
The notation of the presentation as well as the procedures used in
 the NPA method deployed here are identical and hence
the reader is referred to that publication.

\subsection{The NPA pseudopotentials and pair potentials}
The pseudopotentials and pair potentials have been  constructed using a
linear response (LR) approximation.
The KS calculation for the electron states for the NPA in a fluid involves
solving a simple radial equation. The continuum states $\phi_{k,l}(r),
\epsilon_k=k^2/2$, with occupation numbers $f_{kl}$, are evaluated to a
sufficiently large energy cutoff and for an appropriate number of $l$-states
(typically 9 to 39 were found sufficient for the calculations presented here).
 The very high-$k$ contributions are included
by a Thomas-Fermi correction. This leads to an evaluation of the
free-electron density $n_f(r)$, and the free-electron density pileup $\Delta
n'(r)=n_f(r)-\bar{n}$. A part of this pileup is due to the presence of the
cavity potential. This contribution $m(r)$ is evaluated using its linear
response to the electron gas of density $\bar{n}$ using the interacting
electron response $\chi(q,T_e)$. The cavity corrected free-electron pileup
$\Delta n_f(r)=\Delta n'(r)-m(r)$ is used in constructing the electron-ion
pseudopotential as well as the ion-ion pair potential $V_{ii}(r)$ according to
the following equations (in Hartree atomic units), given  for
Fourier-transformed quantities:
\begin{eqnarray}
\label{pseudo.eq}
U_{ei}(k) &=& \Delta n_f(k)/\chi(k,T_e),\\
\label{resp.eq}
\chi(k,T_e)&=&\frac{\chi_0(k,T_e)}{1-V_k(1-G_k)\chi_0(k,T_e)},\\
\label{lfc.eq}
G_k &=& (1-\kappa_0/\kappa)(k/k_{\rm TF});\quad V_k =4\pi/k^2,\\
\label{ktf.eq}
k_{\rm TF}&=&\{4/(\pi \alpha r_s)\}^{1/2};\quad \alpha=(4/9\pi)^{1/3},\\
\label{vii.eq}
 V_{ii}(k) &=& Z^2V_k + |U_{ei}(k)|^2\chi_{ee}(k,T_e).
\end{eqnarray}
Here $\chi_0$ is the finite-$T$ Lindhard function, $V_k$ is the bare Coulomb
potential, and $G_k$ is a local-field correction (LFC). The finite-$T$
compressibility sum rule for electrons is satisfied since $\kappa_0$ and
$\kappa$ are the non-interacting and interacting electron compressibilities
respectively, with  $\kappa$ matched to the $F_{xc}(T)$ used in the KS
calculation. In Eq.~\ref{ktf.eq}, $k_{\rm TF}$ appearing in the LFC is the
Thomas-Fermi wavevector. We use a $G_k$ evaluated at $k\to 0$ for all $k$
instead of the more general $k$-dependent form (e.g., Eq.~50  in
Ref.~\cite{PDWXC}) since the $k$-dispersion in $G_k$ has negligible effect for
the WDMs of this study. The xc-functional is used in the LDA which is efficient
and accurate because the one-center electron density $n(r)$ is smooth compared
to the complex $N$-center electron density used in VASP-type $N$-center DFT
calculations.

Unlike Si$^{4+}$
with a robust core, the  C$^{4+}$ ion produces strong
interactions, especially in the low-$\bar{\rho}$, low-$T$ region.
Stanek  {\it et al}~\cite{Stanek21}  showed that
linear-response (LR)  potentials generated from NPA
for $l$-C at the ``graphite density''
$\rho_{\rm G}\simeq$ 2.267 g/cm$^3$ at low $T$ strongly
over-estimated the first peak of the ion-ion pair distribution
function (PDF) $g(r)$. LR is known to fail for expanded
 metals at low $T$ and here we only study
densities $\bar{\rho}>$ 2.9 g/cm$^3$.

On the other hand, QMD calculations with smaller $N$ lead to
pair-potentials where the long-ranged Friedel oscillations
are not accurately implemented (see Fig.~\ref{Vr.fig},
 and Ref.~\cite{CPP-carb18}).
While this may not be too important in many liquid metals, they
may be crucial to $l$-carbon and kindred materials where a 2$k_F$
subpeak occurs in the structure factor and enforces a high degree of
ordering in $k$-space.

\begin{figure}[t]
\includegraphics[width=0.9\linewidth,keepaspectratio]{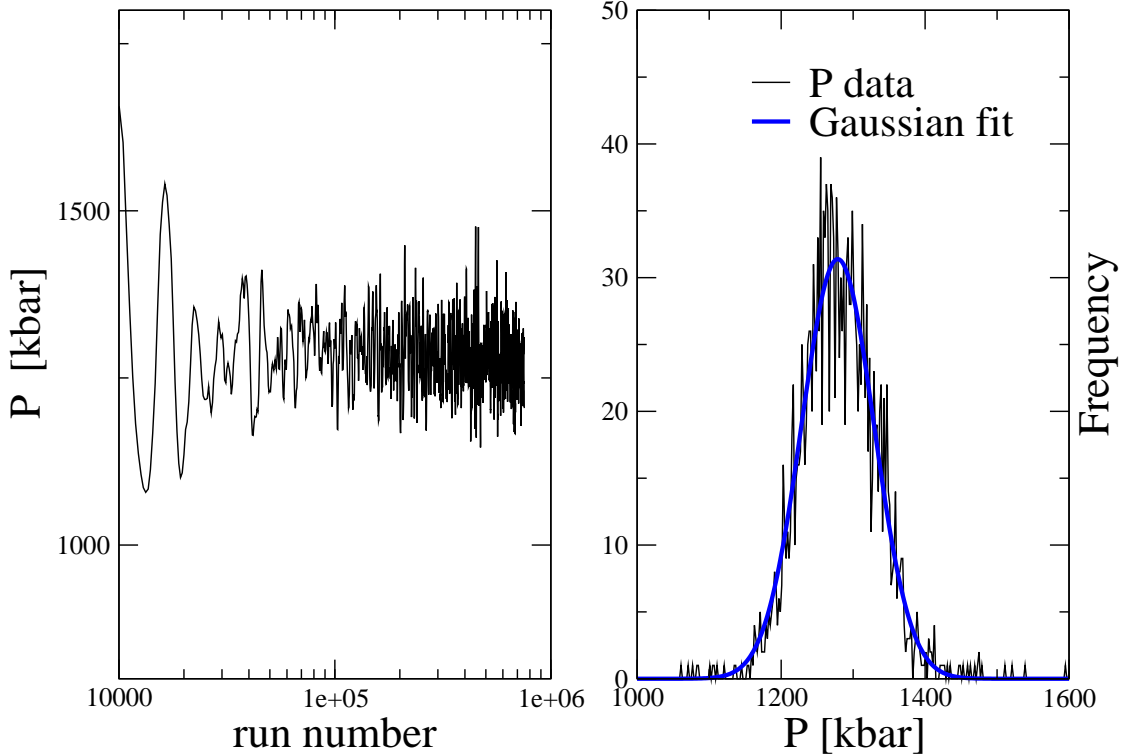}
\caption{
(Color online) (Left Panel) Pressure `readings' at 1 eV
 from a DFT-MD simulation using 108 atoms at 3.1 g/cm$^3$
on the low density side prior to the LPT$_3$. The `run number'
 is an arbitrary index.
(Right panel) The pressure `readings' are fitted to
a Gaussian (unnormalized), and the equilibrium pressure is
taken to be that corresponding to the maximum of the Gaussian.
In the present case a bimodal character is clear and a sum
 of two Gaussians with weights of 0.699 and 0.301
is  appropriate.
\label{P-Gaus.fig}
}
\end{figure}

\subsection{\bf The Gaussian distributions for QMD-SCAN estimates of
 the equilibrium pressure.}
Since the evidence for the LPTs are to be
based on breaks in the predicted pressure,  particular care was used to
determine the equilibrium pressure from the simulation in a consistent,
non-subjective manner. The QMD simulation records a distribution
 of pressures around the thermodynamic mean value, and they form
 a Gaussian distribution for a {\it uniform fluid} at equilibrium.
We expect the Gaussian to be more narrow when the number of
atoms $N$ in the QMD simulation is increased. Hence we show
the distributions for $N$=108 and $N=64$ ins several cases.

Hence the simulation data were fitted to a Gaussian form and the
most probable pressure was determined from the peak of the
Gaussian. Very close to a phase transition, relevant physical quantities
show larger fluctuations. They may not show a unique value and become
 ill-defined.  In two cases, viz., near the LPT$_3$ and
LPT$_{3.7}$, an average over two Gaussians seemed
more appropriate.

\begin{figure}[t]
\includegraphics[width=0.9\linewidth,keepaspectratio]{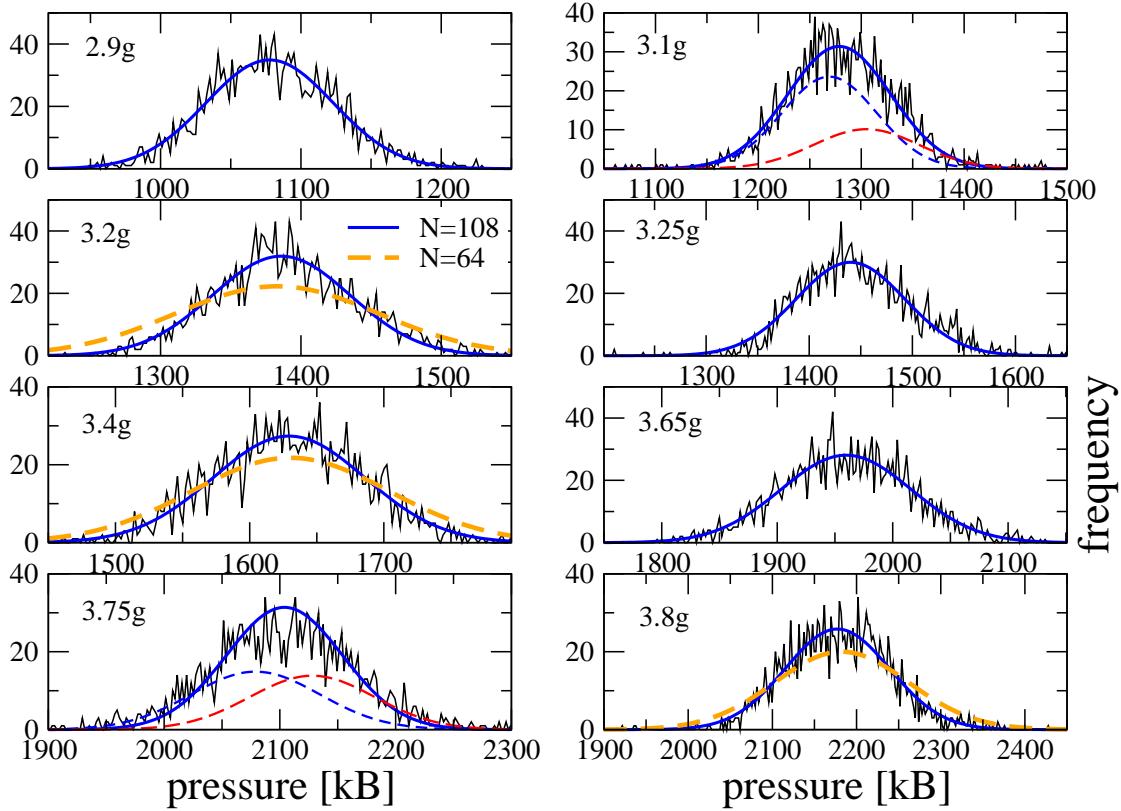}
\caption{
(Color online)  Pressure `distributions' at 1 eV
fitted to Gaussians. The distributions
at $\bar{\rho}$=3.1 g/cm$^3$ (see LPT$_3$), Fig.4 and
 $\bar{\rho}$=3.75 g/cm$^3$
(LPT$_3.7$) are broader and use a two-Gaussian fit.
Although there is a discontinuity in the
QMD-SCAN pressure at 3.42 g/cm$^3$ (LPT$_{3.5}$, the $P$ distribution
at 3.4 g/cm$^3$ shows no clear tendency to a bimodal distribution. The
$N=64$ distributions are also shown for the densities
3.2, 3.4 and 3.8 g/cm$^3$.
 \label{P-Gaus8.fig}
 }
\end{figure}

The pressure evaluated in this manner, using the SCAN xc-functional
is plotted as SCAN-P in Fig.~4(b) of the main text. The width
of the unnormalized Gaussian distribution, $\Delta P$ is used
to characterize the density fluctuations in the system and are displayed
in panel (b) of Fig. 4 of the main text, and labeled SCAN-$\Delta P$.
Using the fit form $f(p)=a\exp\{-w^2(p-p_m)^2\}$ for the frequency
of occurrence of the pressure, $p_m$ is the mean pressure, while $w$
was taken as a measure of the width of the distribution.
The width has been multiplied  by a factor of 3.3 in order to plot it
in the same range as the other curves.\\

It should be appreciated that the pressure, a thermodynamic property,
is effectively undefined exactly at the discontinuity. Close to a
phase transition, if a simulation with a finite number $N$ of particles
is carried out, the width of the data narrows as  $N$ approaches the
thermodynamic limit. The existence of two peaks in the distribution does
not mean that there are `two' thermodynamic pressures, but merely a
property of the smaller $N$ simulation and the effect of density
fluctuations near a phase transition when fluctuations
between two possible fluid states become possible. We have used simulations
at $N=64$ and $N=108$. However, statistical effects scale
only as $1/\sqrt{N}$, and further study is warranted in future investigations.

\begin{figure}[t]
\includegraphics[width=0.95\columnwidth]{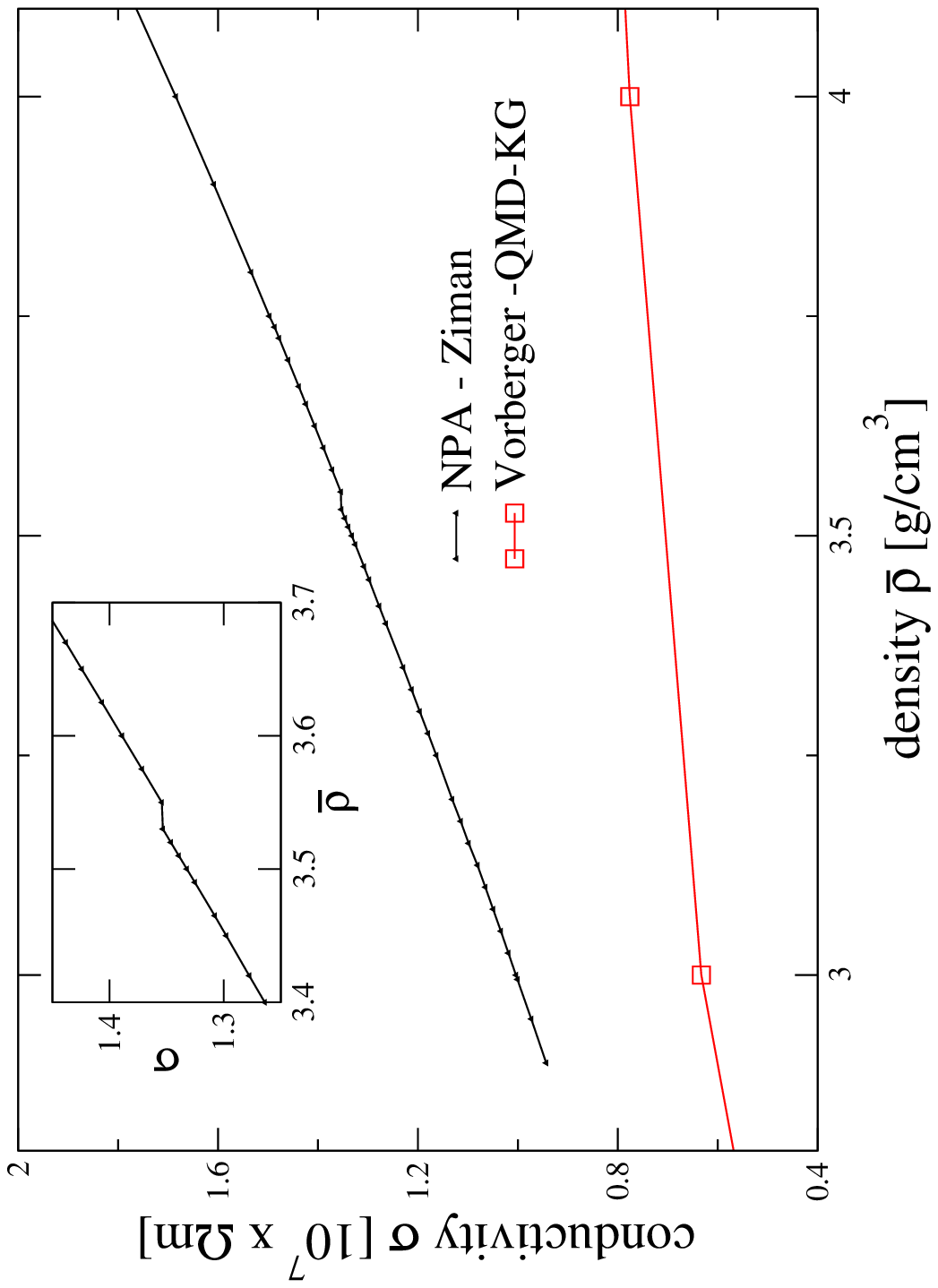}
\caption{
(Color online)
The conductivity calculated using the NPA pseudopotential $U_{ei}(k)$ and
the structure factor $S(k)$ in the Ziman formula is displayed. The conductivity
results reported by Vorberger {\it et al} is also displayed. The slight
discontinuity in $\sigma$ at $\simeq 3.53$ g/cm$^3$ is shown in more detail
in the inset. The discontinuity coincides with the LPT$_{3.5}$. The conductivity
shows no discernible discontinuities at the other two LPTs.
\label{sigma.fig}
}
\end{figure}

\subsection{ The static electrical conductivity $\sigma$}
The static conductivity of $l$-carbon has been a matter of
long-standing concern in regard to the magnetism of giant planets~\cite{Ross81}
and in white dwarfs~\cite{Dufour07}.
 The conductivity usually changes at phase
 transitions because at least one of several physical parameters change
abruptly. These are, e.g.,  the number
of free carriers, the distribution of the ions that cause the
scattering of electrons, and the scattering potential. These
properties are contained in the pseudopotential $U_{ei}(k)$, the structure factor
$S(k)$ and the screening function of the system, when considering the
 static conductivity.

In the present case $\bar{Z}$ remains unchanged at four.
The high value of the Fermi energy $E_F$ of liquid carbon, e.g., 28.8 eV near
 the diamond density of 3.5 g/cm$^3$, implies that at 1 eV the electrons are
 nearly completely degenerate, and hence electron scattering can occur only
 from one edge of the Fermi surface to the
other, with a momentum transfer $q=2k_F$, where $k_F$ is the Fermi momentum. Thus
the quantities  $U_{ei}(2k_F), S(2k_F)$ and the inverse
 dielectric function at $2k_F$ which screens the pseudopotential
are the physical quantities that  determine the
electrical conductivity.

We noted that, due to the strong electron-ion
 scattering at 2$k_F$, the structure factor remains tied to a peak at 2$k_F$.
Thus sharp changes are NOT expected in the electrical conductivity of this
system at liquid-liquid phase transitions.
 However, an extremely weak discontinuity can be seen
 (Fig.~\ref{sigma.fig}) at the LPT$_{3.5}$, i.e.,  at the nominal diamond density.
 Here we also noted a  discontinuity in the pressure calculated via QMD-SCAN,
and via the NPA model, corresponding the reduction of the
form factor of the pseudopotential at 2$k_F$ to zero at this density.
Given the very small magnitude of the discontinuity in
$\sigma$ at LPT$_{3.5}$, it is not surprising that no discontinuities in
 $\sigma$  are seen at
LPT$_3$ and LPT$_{3.7}$ where the QMD pressure isotherm also did not show any
discontinuities.

The electrical conductivity estimate by Vorgerber {\it et al}~\cite{Vorberger20}, using
the PBE-functional and the Kubo-Greenwood (KG) formula is also given, and reports a
lower conductivity. The static conductivity is estimated from the dynamic conductivity
of the KG calculation by an extrapolation to the static limit and averaged over.
The use of a Drude form for this extrapolation is often used in spite of
the importance of transient bonds in these systems.
 It is known that the theoretical predictions and experimental
data when available (e.g., for $l$-silicon) are only in qualitative
 agreement~\cite{cdwSi20} for
these metallic transiently bonded fluids.


\begin{thebibliography}{99}

\bibitem{Hull20}
C. J. Hull,  S. L. Raj, and R. J.  Saykally,
%The liquid state of carbon,
Chemical Physics Letters {\bf 749}, 137341 (2020).

\bibitem{Lazicki21}
A. Lazicki, D. McGonegle, J. R. Rygg, D. G. Braun, D. C. Swift, M. G. Gorman,
 R. F. Smith, P. G. Heighway, A. Higginbotham, M. J. Suggit, D. E. Fratanduono,
 F. Coppari, C. E. Wehrenberg, R. G. Kraus, D. Erskine, J. V. Bernier,
 J. M. McNaney, R. E. Rudd, G. W. Collins, J. H. Eggert and  J. S. Wark,  
%Metastability of diamond ramp-compressed
%to 2 terapascals,
Nature {\bf 589}, 532 (2021).


\bibitem{galli89}
 G. Galli, R. Martin, R. Car,
 and M. Parrinello, Phys.Rev. Lett. {\bf 63}, 988 (1989).
%Carbon: The nature of the liquid state


\bibitem{Benedict14}
Lorin X Benedict, Kevin P Driver, Sebastien Hamel, Burkhard Militzer,
 Tingting Qi, Alfredo A Correa, A Saul, Eric Schwegler,
%Multiphase equation of state for carbon addressing high pressures and temperatures
Phys. Rev. B {\bf 22} 224109 (2014).


\bibitem{Vorberger20}
J. Vorberger, K. U. Plageman, and R. Redmer,
%The structure in warm dense carbon,
High Energy Density Physics {\bf 35}, 100737 (2020).


\bibitem{DWP-Carb90}
M. W. C. Dharma-wardana,and  F. Perrot,  
%Density-functional study of C, Si, and Ge metallic liquids.
Phys. Rev. Lett.,
{\bf 65}, 76 (1990).



\bibitem{GaneshSiLPPT-09}
P. Ganesh, P. and M. Widom, M.
%Liquid-Liquid Transition in Supercooled Silicon Determined by First-Principles
% Simulation,
Phys. Rev. Lett. {\bf 102}, 075701 (2009).


\bibitem{cdwSi20}
M. W. C. Dharma-wardana,
%M. W. C.,
Denis. D. Klug,
and  R. C. Remsing,
%Liquid-liquid Phase Transitions in Silicon,
Phys. Rev. Lett. {\bf 125}, 075702 (2020).




\bibitem{Zong21}
%Zong, H. {\it et al}
H. Zong, V. N. Robinson, A. Hermann, and L. Zhao
%Free electron to electride transition in dense liquid
%potassium,
Nature Physics, {\bf 17}, 955 (2021).

\bibitem{cdw-Carbon10E6-21}
M. W. C. Dharma-wardana,
%Ionization of carbon at 10-100 times the diamond density and in
% the 10$^6$ K temperature range,
Phys. Rev. E {\bf 104}, 015201 (2021).


\bibitem{vanThiel93}
M. V. Thiel, and F. H. Ree,  
%High-pressure liquid-liquid phase change in carbon,
Phys. Rev. B {\bf 48}, 3591 (1993).


 
\bibitem{glosli99}
J. N. Glosli, and   F. H. Ree,
%Liquid-Liquid Phase Transformation in Carbon,
Phys. Rev. Lett. {\bf 82}, 4659 (1999).

\bibitem{PBE96}
J. P. Perdew, K. Burke, K., and  M. Ernzerhof,
%Generalized Gradient Approximation Made Simple,
 Phys. Rev. Lett. {\bf 77}, 3865 (1996).


\bibitem{WuLPT02}
C. J. Wu, J. N. Glosli, G. Galli,
and F. H. Ree,
%Liquid-Liquid Phase Transition in Elemental Carbon:
% A First-Principles Investigation,
Phys. Rev. Lett. {\bf 89}, 135701 (2002).


\bibitem{kraus13}
D. Kraus, D.,
 J. Vorberger, D. O. Gericke, V. Bagnoud, A. Blazevic, W. Cayzac,
 A. Frank, G. Gregori, A. Ortner, A. Otten, F. Roth, G. Schaumann,
 D. Schumacher, K. Siegenthaler, F. Wagner, K. Wunsch, and M. Roth,
%Probing the Complex Ion Structure in Liquid Carbon at 100 GPa,
Phys. Rev. Let.  {\bf 111}, 255501 (2013).


\bibitem{Stanek21}
 Lucas J. Stanek, Raymond C. Clay III, M. W. C. Dharma-wardana,
 Mitchell A. Wood,
 Kristian R. C. Beckwith, and Michael S. Murillo,
Phys. Plasmas {\bf 28}, 032706 (2021).

\bibitem{VASP}
G. Kresse, G. and  Furthm\"{u}ller, J.
%Efficient iterative schemes for ab initio total-energy calculations using
% a plane-wave basis set.
Phys. Rev. B \textbf{54}, 11169 (1996).

\bibitem{ABINIT}
%Gonze, X. and Lee, C.
%ABINIT: First-principles approach to material and nanosystem properties,
%Computer Phys. Commun. \textbf{180}, 2582-2615 (2009).
% The Abinit Project: Impact, Environment and Recent Developments.
X.Gonze, B. Amadon, G. Antonius, F. Arnardi, Lucas Baguet,
Jean-Michel Beuken, Jordan Bieder, F.  Bottin,
Johann Bouchet, Eric Bousquet,
{\it et al},
Computer Physics Communications, {\bf 248}, 107042 (2020).
%https://doi.org/10.1016/j.cpc.2019.107042.





\bibitem{Remsing17}
R. C. Remsing, M. L. Klein, and  J. Sun,
%Dependence of the structure and dynamics of liquid silicon on the choice of
%density functional approximation,
Physical Review B {\bf 96} 024203 (2017).




\bibitem{SCAN13}
J. Sun, B. Xiao, Y. Fang, R. Haunschild, P. Hao, A. Ruzsinszky,
G. I. Csonka, G. E. Scuseria, and J. P. Perdew,
Phys. Rev. Lett.
{\bf 111}, 106401 (2013).


\bibitem{eos95}
F. Perrot, and M. W. C. Dharma-wardana,
%Equation of state and transport properties of an interacting multispecies plasma:
% Application to a multiply ionized Al plasma,
Phys. Rev. E. {\bf 52}, 5352 (1995).

\bibitem{Hungary16}
M. W. C. Dharma-wardana,
% Current Issues in Finite-$T$ Density-Functional Theory
% and Warm-Correlated Matter. 50$^{th}$ anniversary of Kohn-Sham theory.
Proceedings of the Conference on Density Functional Theory, Debrecen, 2015.
Edited by K. Schwarz and A. Nagy.
Computation  {\bf 4} (2), 16; 2016,







\bibitem{DSF18}
L Harbour, and G. D. F\"orster,  M. W. C. Dharma-wardana and
 Laurent J.Lewis,
%Ion-ion dynamic structure factor, acoustic modes, and equation of state of
%two-temperature warm dense aluminum.
Physical review E {\bf 97},043210 (2018).




\bibitem{whitley15}
H. D. Whitley,
D. M. Sanchez , S. Hamel , A. A. Correa,
and L. X. Benedict, Contrib. Plasma Phys. {\bf 55}, 390 (2015).





\bibitem{DWP82}
M. W. C. Dharma-wardana, and  F. Perrot,
%Density functional theory of hydrogen plasmas,
Phys. Rev. A {\bf 26}, 2096  (1982).


\bibitem{Aers-CDW-Gibb86}
G.C. Aers, M.W.C. Dharma-wardana and M. Gibb,
Phys. Rev. B {\bf 33}, 4307 (1986).  [NRCC 25290]
%Liquid Ge:  A Strongly
%Coupled Liquid with Negligible Bridge Contributions to the Structure Factor.



\bibitem{PDWXC}
F. Perrot, and M. W. C. Dharma-wardana,
%Spin-polarized electron liquid at arbitrary temperatures:
%Exchange-correlation energies,
%electron-distribution functions, and the static response functions,
Phys. Rev. B {\bf 62}, 16536 (2000);
{\it Erratum: } {\bf 67}, 79901 (2003).


\bibitem{Dornheim18}
%T. Dornheim, {\it et al},
T. Dornheim, G. Groth, G. and B. Bonitz,
%The uniform electron gas at warm dense matter conditions,
Physics Reports, {\bf 744}, 1-86  (2018).



\bibitem{cdw-N-rep19}
M. W. C. Dharma-wardana,
%Simplification of the electron-ion many-body problem: N-representability
%of pair densities obtained via a classical map for the electrons,
Phys. Rev. B {\bf 100}, 155143 (2019).
%DOI: 10.1103/PhysRevB.100.155143



\bibitem{Fauss21}
G. Faussurier, C. Blancard,  and  M. Bethkenhagen,
C%arbon ionization from a quantum average-atom model up to gigabar pressures.
Phys. Rev. E {\bf 104}, 025209 (2021).


\bibitem{Anta00}
J. A. Anta and A. A. Louis,
%Probing ion-ion and electron-ion correlations in liquid metals within
%the quantum hypernetted chain approximation,
Phys. Rev. B {\bf 61}, 11400 (2000).   




\bibitem{Murillo13}
M. S. Murillo,     
 J. Weisheit, S. B. Hansen, and M. W. C. Dharma-wardana,
Phys. Rev. E {\bf 87}, 063113 (2013).


\bibitem{Rosznyai2008}
%Electron scattering in hot/warm plasmas,
Balas F. Rosznyai, High Energy Density Physics {\bf 4} 64 (2008)


\bibitem{StarretHam14}
C. E. Starrett, D. Saumon, J. Daligault, and S. Hamel,
%Integral equation model for warm and hot dense mixtures,
Phys. Rev. E {\bf 90}, 033110 (2014).

\bibitem{YongHou-AA-17}
Yong Hou, Yongsheng Fu, Richard Bredow, Dongdong Kang, Ronald Redmer,
 Jianmin Yuan,
%Average-atom model for two-temperature states and ionic transport properties of aluminum
% in the warm dense matter regime,
High Energy density Physics {\bf 22}  21 (2017).


\bibitem{Pe-Be}
F. Perrot,
%Ion-ion interaction and equation of state of a dense plasma: Application to beryllium,
Phys. Rev. E {\bf 47}, 570 (1993).


\bibitem{Ross81}
M. Ross,  
%The ice layer in Uranus and Neptune-diamonds in the sky?
Nature {\bf 292} 475 (1981).

\bibitem{Dufour07}
 P. Dufour, J. Liebert, G. Fontaine, N. Behara,
%White dwarf stars with carbon amospheres,
Nature {\bf 450},  522 (2007).



\bibitem{Baldereshi73}

A. Baldereschi,
%Mean-value point in the Brillouin zone,
Phys. Rev. B, {\bf 7}, 5212 (1973).

\bibitem{Chihara78}
J. Chihara, Progr. Theor. Phys. Jpn. {\bf 59}, 1085 (1978).

\bibitem{EvansDFT79}
R. Evans,
%The nature of the liquid-vapour interface and other topics in the statistical mechanics of non-uniform, classical fluids,
Adv. Phys. {\bf 28}, 143 (1979).
% cklassical DFT for an interface


\bibitem{LFA83}
F. Lado, S. M. Foiles,  and N. W. Ashcroft,
%Solutions of the reference-hypernetted-chain equation with minimized free energy,
Phys. Rev. {\bf A 26}, 2374 (1983).




\bibitem{cdw-pop21}
M. W. C. Dharma-wardana,
%Simple pair-potentials and pseudo-potentials for warm-dense matter and general applications,
Physics of Plasmas {\bf 28}, 052108 (2021).

\bibitem{CPP-carb18}
M. W. C. Dharma-wardana,
%Theory of complex fluids in the warm dense matter regime and application to
% an unusual phase transition in liquid carbon,
Contrib. Plasma Phys. {\bf 58}, 128-142 (2018).





\end{thebibliography}
\end{document}